\documentclass[a4paper,11pt]{article}
\usepackage{jheppub} 

\usepackage[utf8]{inputenc}
\usepackage[colorlinks=true,citecolor=blue,linkcolor=blue]{hyperref}
\usepackage[normalem]{ulem}
\usepackage{amsmath,amssymb}
\usepackage{epsfig}
\usepackage{graphicx} 
\usepackage{placeins} 
\usepackage{url}
\usepackage{color}
\usepackage{slashed}
\usepackage{multirow}
\usepackage{enumerate}
\usepackage{placeins}
\usepackage[dvipsnames]{xcolor}
\usepackage{epstopdf}
\usepackage{soul}
\usepackage{tikz}
\usepackage[capitalise, english]{cleveref}
\usepackage{siunitx}
\usepackage{xspace}
\usepackage{subcaption}
\usepackage{array}
\usetikzlibrary{trees}
\usepackage{comment}
\usetikzlibrary{decorations.pathmorphing}
\usetikzlibrary{decorations.markings}
\usetikzlibrary{calc,tikzmark,fit,shapes.geometric,matrix,decorations.markings,arrows.meta,decorations.pathmorphing,patterns,positioning,snakes}

\tikzset
  {midarrow/.style={decoration={markings,mark=at position 0.5 with
     {\arrow[thin,xshift=2pt]{Triangle[length=4pt,#1]}}},postaction={decorate}}
  }

\tikzset{
proton/.style = {circle, draw=black, thin, fill=black!20!white, minimum size=#1,
              inner sep=0pt, outer sep=0pt},
proton/.default = 6pt % size of the circle diameter 
}

\tikzset{
blob/.style = {circle, draw=black, thin, preaction={fill, black!20!white}, pattern=north east lines, minimum size=#1,
              inner sep=0pt, outer sep=0pt},
blob/.default = 6pt % size of the circle diameter 
}

\tikzset{
wc/.style = {circle, fill, minimum size=#1,
              inner sep=0pt, outer sep=0pt},
wc/.default = 4pt % size of the circle diameter 
}

\tikzset{vector/.style={decorate, decoration=snake}}

\newcommand\myshade{80}
\colorlet{mylinkcolor}{ForestGreen}
\colorlet{mycitecolor}{Red}
\colorlet{myurlcolor}{violet}

\hypersetup{
  linkcolor  = mylinkcolor!\myshade!black,
  citecolor  = mycitecolor!\myshade!black,
  urlcolor   = myurlcolor!\myshade!black,
  colorlinks = true
}
\definecolor{jblue}{RGB}{20,50,100}
\definecolor{npurple}{RGB} {153, 51, 204}
\definecolor{wred}{RGB}{217,0,56}
\definecolor{white}{RGB}{255,255,255}

\definecolor{korange}{RGB}{235, 80,  43}
\definecolor{korange2}{RGB}{245, 100,  63}
\definecolor{kyelloworange}{RGB}{255, 210,  110}
\definecolor{kyelloworange2}{RGB}{240, 170,  90}
\definecolor{kred}{RGB}{204,  102, 153}
\definecolor{kpurple}{RGB}{153,  61, 190}
\definecolor{kpurplelight}{RGB}{213,  161, 230}

\allowdisplaybreaks

\def\hyp{\mathsf{y}}
\def\nn{\nonumber\\ }

\newcommand{\eps}{\varepsilon}

\newcommand{\cC}{\mathcal{C}}

\newcommand{\cO}{\mathcal{O}}

\newcommand{\eg}{\emph{e.g.}~}
\newcommand{\ie}{\emph{i.e.}~}
\newcommand{\EW}{\mu_{\rm EW}}

\makeatletter 
\gdef\@fpheader{}
\makeatother
\begin{document}

%=============================================================================

\title{Non-universal probes of composite Higgs models: New bounds and prospects for FCC-ee}

\author[a]{Ben A. Stefanek}
\emailAdd{benjamin.stefanek@kcl.ac.uk}

\affiliation[a]{Physics Department, King’s College London, Strand, London, WC2R 2LS, United Kingdom}

\date{\today}

\preprint{KCL-PH-TH/2024-43}

%=============================================================================

\abstract{
We study the leading loop-level phenomenology of composite Higgs models via the effective field theory of a strongly interacting light Higgs and top quark (SILH+TQ). We systematically analyze the renormalization group evolution (RGE) of tree-generated operators in the SILH+TQ scenario, finding large mixings of flavor non-universal operators into those affecting electroweak precision observables. We show that these model-independent RG contributions are more important than typical estimates for finite matching terms. Flavor non-universal effects are completely captured by examining three options for the top mixing: fully composite $q_L^3$, equal compositeness, and fully composite $t_R$. In the most phenomenologically viable case of a fully composite $t_R$, we show that the strongest bound on the natural parameter space comes from a 2-loop double-log contribution of the 4-top operator $\cO_{tt} = (\bar t_R \gamma_\mu t_R)(\bar t_R \gamma^\mu t_R)$ into the Peskin–Takeuchi $T$ parameter. In general, we find that this 2-loop effect allows existing electroweak precision data to give better constraints on 4-top operators than high-energy probes from top production at the LHC. Independent of the top mixing, we find that a future tera-$Z$ machine such as FCC-ee has the potential to probe composite Higgs models up to a scale of $m_* \gtrsim 25$ TeV, and test the naturalness of the electroweak scale at the $\lesssim 10^{-4}$ level.
}

\maketitle

%-----------------------------------------------------------------------------
\section{Introduction}
Composite Higgs models and their warped five-dimensional (5D) duals remain one of the best motivated solutions to the electroweak (EW) hierarchy problem~\cite{Randall:1999ee}. Another compelling feature of these models was the opportunity to explain the hierarchies observed in the Standard Model (SM) Yukawa couplings without imposing any flavor symmetries. In this ``flavor anarchy" paradigm, the Yukawa matrices are assumed to be random at short distances, and hierarchies are generated dynamically in the infrared (IR) via renormalization group evolution (RGE)~\cite{Kaplan:1991dc,Nelson:2000sn}. The dual description of this picture is the quasi-localization of different fermion generations at different points in the warped 5D bulk~\cite{Grossman:1999ra,Gherghetta:2000qt,Huber:2000ie}. 

Historically, however, these models have faced stringent constraints due to their predictions of CP violation, flavor-changing neutral currents (FCNC), and large violations of custodial symmetry, calling their naturalness into question. Though the so-called RS-GIM mechanism provides significant flavor protection, it has been known for a long time that it is not enough to reach the TeV scale~\cite{Burdman:2002gr,Burdman:2003nt,Huber:2003tu,Agashe:2004ay,Agashe:2004cp,Csaki:2008zd,Isidori:2010kg}. This was recently reanalyzed in detail in~\cite{Glioti:2024hye}, with the conclusion that anarchic partial compositeness in the quark sector is incompatible with new physics (NP) below about 20 TeV. One possibility is to give up on an explanation of the flavor hierarchies and assume minimal flavor violation~\cite{DAmbrosio:2002vsn}, which is the hypothesis of flavor universality broken by the minimal set of spurions: the SM Yukawa couplings themselves. However, the ATLAS and CMS collaborations have set bounds on flavor-universal NP of order of 5-10 TeV~\cite{CMS:2021ctt,ATLAS:2023fod,ATLAS:2017eqx,ATLAS:2020yat}, so low-scale models now face problems with flavor-conserving observables. 

As known already for some time in the flavor community, this necessitates the conclusion that less than maximal flavor symmetries are in fact optimal for achieving the lowest NP scale. In particular what is needed is NP coupling universally to the light families, while it may have a different coupling to the third generation, a scenario described by $U(2)^n$ flavor symmetries~\cite{Feldmann:2008ja,Kagan:2009bn,Barbieri:2011ci,Barbieri:2012uh,Blankenburg:2012nx,Barbieri:2012tu,Barbieri:2012bh,Fuentes-Martin:2019mun,Greljo:2022cah}. Invoking them opens up the possibility to relax LHC bounds while maintaining sufficient protection from the strongest FCNC constraints, if the symmetry is minimally broken~\cite{Allwicher:2023shc}. Dynamical origins for $U(2)^n$ symmetries have been studied in great detail in recent years, culminating in two key directions. The first is a revitalization of the old idea of flavor non-universal gauge symmetries, into which the SM gauge group is vertically embedded. Now called flavor deconstruction, this idea can realize $U(2)^n$ symmetries as an accident of the extended, flavor non-universal gauge sector~\cite{Bordone:2017bld,Greljo:2018tuh,Fuentes-Martin:2020bnh,Davighi:2022fer,FernandezNavarro:2022gst,Davighi:2022bqf,FernandezNavarro:2023rhv,Davighi:2023iks,Davighi:2023evx,FernandezNavarro:2023hrf,Fuentes-Martin:2024fpx,Greljo:2024ovt}.

The second direction is the possibility of realizing $U(2)^n$ as a dynamical low-energy accident, essentially because any NP distinguishing the light flavors is very heavy. This is naturally implemented in the context of a strongly-coupled sector with multiple mass gaps at hierarchical scales~\cite{Panico:2016ull}. The dual description of such a theory corresponds to a warped extra dimension with multiple branes~\cite{Fuentes-Martin:2020pww,Fuentes-Martin:2022xnb,Stefanek:2022doo,Greljo:2019xan}. While the size of the top Yukawa clearly indicates an IR realization à la partial compositeness, the key insight here is that the light fermion Yukawas may originate instead from strong sector operators that decouple when their constituents acquire mass at scales high enough to avoid flavor bounds, generating higher-dimensional bilinear interactions.
Integrating out the heavy NP associated with the bilinears leads to an accidental $U(2)^n$ symmetry at low energies, with leading breaking by a doublet spurion of $U(2)_q$ responsible for generating light-heavy CKM mixing in the IR.\footnote{This last point follows automatically if the light generation right-handed fermions are elementary from the IR point of view (i.e. in the 5D language, they should be fully localized in their respective branes).}

While it is clear that additional flavor structure is needed on top of the anarchic scenario in order to have NP at the TeV scale, we will remain mostly agnostic to it here and focus on the one irreducible aspect of the IR physics common to all composite Higgs models- the fact that $q_L^3$ and $t_R$ must have large mixings with the strong sector in order to obtain $y_t \simeq 1$. Since all other Yukawa couplings are small, the fermionic mixing interactions with the strong sector could have an approximate quark flavor symmetry as large as $U(2)_q \times U(2)_u \times U(3)_d$. This is what we will assume here, as it will allow us to separate out the top quark in a consistent manner~\cite{Redi:2012uj}, while neglecting all other mixing effects that are phenomenologically subleading. 

On top of this, we will always assume that the strong sector respects CP and custodial symmetry. In this case, the leading deviations in EW precision observables are expected to come from tree-level corrections to the $Z  b_L \bar b_L$ vertex~\cite{Agashe:2005dk}, due to the fact that $q_L^3$ must have significant degree of compositeness in order to obtain the top Yukawa. This pushes the NP scale above 4 TeV, unless a custodial $P_{LR}$ symmetry is invoked that can simultaneously protect the $Z b_L \bar  b_L$ and $Z t_R \bar t_R$ couplings at tree-level~\cite{Agashe:2006at}.\footnote{We note that $P_{LR}$ can be realized rather naturally, as it just corresponds to a particular choice of how to embed the SM fermions into representations of the strong sector global symmetry.} If one does all that, the conventional wisdom is that the leading constraints are then flavor-universal ones, namely Higgs coupling modifications~\cite{ATLAS:2022vkf,CMS:2022dwd} and corrections to oblique EW precision observables such as the Peskin–Takeuchi $S$ parameter~\cite{Peskin:1990zt}. The vast majority of attention in the composite Higgs literature has therefore been focused on flavor-universal probes.

Our intent here is to focus instead on how these symmetry-protected models can be probed via irreducible flavor non-universal effects associated with a strongly-interacting top quark (SITQ), namely operators built from $q_L^3$ and $t_R$. We do not claim to be the first to have ever considered this, so let us comment on what has been done and emphasize what is new. One unavoidable aspect in models where the top Yukawa comes from partial compositeness is the existence of sizable 4-top operators. At the moment, the most natural scenario is the limit of a fully composite $t_R$, as it avoids the strongest constraints from flavor-violating operators involving $q_L^3$~\cite{Panico:2012uw,Pappadopulo:2013vca}. In this case, one expects unsuppressed $\cO_{tt} = (\bar t_R \gamma_\mu t_R)(\bar t_R \gamma^\mu t_R)$ operators. The common assumption is that the best bounds on these operators come from high-energy processes, since 4-fermion contact interactions grow as $E^2$. Indeed, LHC bounds were studied in~\cite{Hartland:2019bjb,Ethier:2021bye,Banelli:2020iau,Degrande:2020evl,Aoude:2022deh,Alasfar:2022zyr,Degrande:2024mbg}, while bounds from probing $ee\rightarrow tt$ contact interactions at future linear lepton colliders were studied in~\cite{Durieux:2018ekg,Banelli:2020iau}. Unfortunately, LHC bounds on 4-top operators are not stronger than the leading flavor-universal constraints, while the constraints coming from $ee\rightarrow tt$ are only stronger than universal ones at CLIC, assuming 3 TeV center-of-mass energies.

At the lower energies accessible with circular $e^+ e^-$ machines, it is usually assumed that Higgs compositeness is more likely to manifest itself through universal effects~\cite{Thamm:2015zwa,Durieux:2018ekg,deBlas:2019rxi}. Here we challenge this statement by systematically analyzing the 1-loop RGE of tree-level, flavor non-universal operators generated at the compositeness scale $\Lambda$ and evolved down to the EW scale $\EW$. 
Unlike finite parts of the UV matching frequently estimated in the literature, these RG contributions are model independent. The dominant contributions come from top quark loops and behave as $[y_t^2 N_c \log(\Lambda^2/\EW^2)/(16\pi^2)]^n$. Despite being loop effects, the $y_t^2 N_c$ scaling, $\Lambda \gtrsim $ TeV, and the precision at which EW observables are measured means that they are nonetheless highly relevant.

The only assumptions we make are standard ones, namely i) the strong sector is well described by one coupling $g_*$ and one mass scale $m_*$, ii) a mass gap exists between the EW scale and the scale $\Lambda \sim m_*$ where all composite states can be integrated out, and iii) the top Yukawa is realized via partial compositeness. The second point is justified experimentally by LHC searches for top partners and other composite resonances, yielding $\Lambda \gtrsim$ 1-2 TeV~\cite{ATLAS:2018ziw,CMS:2022fck,ATLAS:2023bfh,ATLAS:2024fdw,ATLAS:2024gyc,ATLAS:2023pja,CMS:2024bni}. The last point is a fundamental feature of all composite Higgs models that we are aware of, including models where the top Yukawa itself is generated via bilinear interactions. The basic reason is that bilinears must have dimension $\geq 5$, so they should be effectively ``UV completed" already by partial compositeness at the scale $\Lambda$~\cite{Panico:2016ull}.

Under these mild assumptions, we can consider a subset of the SM effective field theory (SMEFT) consisting of flavor-universal operators from a strongly-interacting light Higgs (SILH), together with flavor non-universal operators due to a SITQ. 
By considering the resummed 1-loop RGE of this EFT, we show that the leading bound on the fully composite $t_R$ scenario comes from a 2-loop double-log contribution of $\cO_{tt}$ to the operator $\cO_{HD} = |H^\dagger D_\mu H|^2$ that is proportional to the $T$ parameter, constrained at the per-mille level by LEP data. We also show that non-universal RG effects are currently more important than standard estimates for finite matching contributions unless $q_L^3$ is mostly composite, while for a future tera-$Z$ machine they will dominate no matter how one chooses to mix $q_L^3$ and $t_R$. Some of the first leading-log (LL) RG effects have been pointed out before~\cite{Pomarol:2008bh,Grojean:2013qca,Elias-Miro:2013mua,Bellazzini:2014yua,Panico:2015jxa,Feruglio:2017rjo,Banelli:2020iau,Dawson:2022bxd,Allwicher:2023aql,Allwicher:2023shc,Garosi:2023yxg,Bellafronte:2023amz}, while the 2-loop double-log results were first pointed out by us in~\cite{Allwicher:2023aql} as well as later in~\cite{Garosi:2023yxg,Allwicher:2023shc}, but are completely new in the composite Higgs context. To the best of our knowledge, a systematic analysis considering all 1-loop RG effects in a general SILH+TQ EFT framework is missing in the literature, which is precisely the goal of this paper. For example, these RG effects were not considered in the recent update of Ref.~\cite{Glioti:2024hye}. 

In this work, we analyze flavor-universal constraints from EW precision data taking RGE into account, as well as flavor non-universal ones due to a SITQ. The top Yukawa constraint $y_t \simeq g_* \eps_L \eps_R$ means that only one new mixing parameter is introduced. We deal with this by considering three mixing scenarios: i) $q_L^3$ fully composite, ii) equal compositeness of $q_L^3$ and $t_R$, and iii) $t_R$ fully composite. The complementarity of the universal and non-universal constraints to probe different regions of the $(m_*,g_*)$ parameter space will be highlighted, \eg we show that flavor non-universal RGE effects are most important in the $g_*\gtrsim 1$ region. Interestingly, this is also the most natural parameter space since $\langle H \rangle \sim f = m_*/g_*$. We also compare our results to current and future bounds from LHC searches and the leading flavor constraints from $B$ physics. Finally, we project that a future circular $e^+ e^-$ collider such as FCC-ee featuring a $Z$-pole run producing $O(10^{12})$ $Z$ bosons together with a precision measurement of $m_W$ on the $WW$ threshold can set a bound $m_* \gtrsim 25$ TeV on the compositeness scale, independent of the mixing.

The paper is organized as follows. In~\S~\ref{sect:setup}
we give the details of our setup and EFT framework that we employ featuring a SILH+TQ. Next, in~\S~\ref{sect:singleOps} we discuss the details of our EW likelihood and use it to obtain current and future bounds on the Wilson coefficients (WCs) of our EFT. In~\S~\ref{sec:RGEeffects}, we break down the dominant RG effects in detail, giving analytic formulas. We also compare bounds on 4-top operators from EW precision data with those from top production at the LHC. Our main results specific to composite Higgs models are given in~\S~\ref{sect:CHmodels}.
Finally, we discuss and summarize our results in the Conclusions of~\S~\ref{sec:conclusions}.

%----------------------------------------------------------------------------
\section{Setup and EFT framework}
\label{sect:setup}
%----------------------------------------------------------------------------
As is standard in the composite Higgs literature~\cite{Panico:2015jxa}, we assume that at high energies there are two sectors: 1) an elementary sector $\mathcal{L}_{{\rm SM}'}$ consisting of the SM gauge sector and elementary fermions $\psi$, as well as 2) a strongly-coupled sector $\mathcal{L}_{\rm strong}$. These two sectors are coupled via the weak gauging of the SM subgroup of the global symmetry of the strong sector as well as fermionic mixing between them. The high-energy Lagrangian thus takes the schematic form
\begin{equation}
\mathcal{L}_{\rm UV} =\mathcal{L}_{{\rm SM}'}+\mathcal{L}_{\rm strong}+ g A_\mu^{\rm SM} J^\mu_{\rm strong} + \mathcal{L}_{\rm mix}(\psi, \cO_\psi) \,,
\end{equation}
where $g$ is a SM gauge coupling and $\mathcal{L}_{\rm mix}$ consists of mixing terms between the elementary fermions $\psi$ and composite fermionic operators ${\cal O}_\psi$. The usual assumption for $\mathcal{L}_{\rm mix}$ is linear mixings of the form $\bar \psi \boldsymbol{\lambda}_\psi \mathcal{O_\psi}$, for all elementary fermions $\psi$. This is the partial compositeness paradigm which, as discussed in the introduction, leads to severe flavor constraints if the matrix $\boldsymbol{\lambda}_\psi$ has an anarchic structure.

Since we are interested in probing models that allow for the lowest NP scale,  we opt for an approach where we assume that $\mathcal{L}_{\rm mix}$ respects an approximate $\mathcal{G}_{\rm mix} = U(2)_q \times U(2)_u \times U(3)_d$ quark flavor symmetry, the maximum symmetry that is well-respected by the SM Yukawas. We further assume a minimal breaking of this symmetry (as is necessary to generate CKM mixing) via a doublet spurion of $U(2)_q$ that we call $\lambda_q^{i}$, where $i=1,2$ is a light generation flavor index. The mixing Lagrangian can then be written as~\cite{Grojean:2013qca}
\begin{equation}
\mathcal{L}_{\rm mix} =  \lambda_L \bar q_L^3 {\cal O}_L + \lambda_R \bar t_R {\cal O}_R + \lambda_q^{i} \bar q_L^i {\cal O}_q + \mathcal{L}_{\rm light}\,,
\label{eq:Lmix}
\end{equation}
where $\mathcal{L}_{\rm light}$ indicates suppressed contributions responsible for the generation of the light fermion masses and mixings. This structure could come from partial compositeness controlled by appropriate flavor symmetries~\cite{Glioti:2024hye}, or dynamically via bilinear operators that generate the light fermion Yukawas deeper in the UV~\cite{Panico:2016ull,Fuentes-Martin:2020pww,Fuentes-Martin:2022xnb,Stefanek:2022doo}. Since the only important point for us is to isolate the top quark in a consistent manner, we simply assume this structure while remaining agnostic to its origin. We emphasize that our assumption for the breaking of $\mathcal{G}_{\rm mix}$ is consistent if we are after a lower bound on the NP scale, since less minimal breakings generically lead to stronger flavor bounds.
After integrating out all composite states except for the Higgs and the top quark, the leading low-energy effective Lagrangian takes the form~\cite{Giudice:2007fh,Contino:2013kra}
\begin{equation}
\mathcal{L}_{\rm EFT}=\mathcal{L}_{{\rm SM}'}+\frac{m_*^4}{g_*^2}\widehat{\mathcal{L}}_{\rm EFT}\left(\frac{g_*H}{m_*},\frac{D_\mu}{m_*},\frac{gF_{\mu\nu}}{m_*^2},\frac{\lambda_{L} \bar q_L^3}{m_*^{3/2}}, \frac{\lambda_{R} \bar t_R}{m_*^{3/2}}, \frac{\lambda_{q}^i \bar q_L^i}{m_*^{3/2}} , \frac{g_*^2}{16\pi^2 }, \frac{g}{16\pi^2 }\right) \,,
\label{eq:scalingEFT}
\end{equation}
where $F_{\mu\nu}$ is any SM gauge field strength tensor. This gives us scaling rules for how fields and derivatives in the low-energy Lagrangian should appear accompanied by composite Higgs model parameters, \emph{e.g.} all covariant derivatives $D_\mu$ must appear suppressed by $1/m_*$, while fermions come with a factor $\lambda_{L,R,q}/m_*^{3/2}$.
According to these rules, the leading Yukawa couplings have the form
\begin{equation}
\mathcal{L}_{\rm Yuk} \simeq \frac{\lambda_L \lambda_R^*}{g_*} \bar q_L^3 \tilde H t_R + \frac{\lambda_q^{i}\lambda_R^*}{g_*} \bar q_L^i \tilde H t_R + \dots \,,
\label{eq:YukEFT}
\end{equation}
so we identify $y_{t} \simeq \lambda_L \lambda_{R}^* / g_* $. The parameter $\lambda_q^{i}$ giving the leading breaking of $U(2)_q$ is responsible for generating light-heavy CKM mixing, namely $V_{td}$ and $V_{ts}$. It will also control all flavor-violating effects, as we will see in~\S~\ref{sec:FVops}.

It is convenient to work in an EFT basis which allows for a transparent parameter power counting of flavor universal as well as non-universal effects. Widely used in the literature to capture flavor-universal effects is the strongly-interacting light Higgs (SILH) basis~\cite{Giudice:2007fh,Contino:2013kra}, which is our starting point. We define the effective Lagrangian
\begin{equation}
\mathcal{L}_{\rm EFT} = \mathcal{L}_{\rm SM} ^{(d=4)} + \sum_{i} \cC_i \cO_i \,,
\label{eq:pureEFT}
\end{equation}
where the sum runs over all dimension-6 operators $\cO_i$ given in~\cref{tab:UniversalOps,tab:nonUniversalOps}. We note that our basis defined in this way is intentionally redundant, \eg using the gauge field equations of motion, one can see that some of the flavor universal operators in~\cref{tab:UniversalOps} can be mapped to flavor-universal contributions to operators of the form $( i H^\dagger  \overleftrightarrow {D^\mu} H)(\bar \psi \gamma_\mu \psi)$ as well as 4-fermion operators, overlapping with the operators in~\cref{tab:nonUniversalOps}. \emph{The crucial point, however, is that these universal effects will come with a different power counting than the non-universal ones.} As we will see, this point is obscured if one starts from the standard Warsaw basis~\cite{Grzadkowski:2010es}, while it is transparent in the redundant, non-universal SILH+TQ basis that we employ here. Once this power counting is established, we convert to the Warsaw basis for RGE and computation of observables. The full map from the SILH+TQ basis to the Warsaw basis is given in~\cref{app:A}.

We will now use the scaling rules derived from~\cref{eq:scalingEFT} to match the Wilson coefficients in~\cref{eq:pureEFT} to composite Higgs model parameters. In addition to assuming that the strong sector respects CP and custodial symmetry, we also assume it is minimally coupled, as is the case for all known composite Higgs models with $g_* < 4\pi$~\cite{Giudice:2007fh}.

\subsection{Flavor universal operators from a strongly-interacting light Higgs}
\begin{table}
\centering
\begingroup
\renewcommand*{\arraystretch}{1.4}
\begin{tabular}{|c|c|}
 \hline
 \multicolumn{2}{|c|}{\textbf{Flavor universal bosonic operators}} \\ \hline \hline
 $\mathcal{O}_H = \frac{1}{2}\partial_\mu (H^{\dagger} H )\partial^\mu (H^{\dagger} H )$ & $\mathcal{O}_T = \frac{1}{2} ( H^{\dagger} \overleftrightarrow{D}_{\mu}H ) ( H^{\dagger} \overleftrightarrow{D}^{\mu}H )$ \\ \hline
 $\mathcal{O}_W = i \frac{g_2}{2}( H^{\dagger} \overleftrightarrow{D}_{\mu}^I H ) D_{\nu} W^{I\, \mu \nu}$ & $\mathcal{O}_B = i \frac{g_1}{2}( H^{\dagger} \overleftrightarrow{D}_{\mu} H ) \partial_{\nu} B^{\mu \nu}$ \\ \hline
 $\mathcal{O}_{2W} = -\frac{g_2^2}{2} (D^{\mu}W_{\mu \nu}^I) (D_{\rho}W^{I \rho \nu})$ & $\mathcal{O}_{2B} = -\frac{g_1^2}{2} (\partial^{\mu}B_{\mu \nu}) (\partial_{\rho}B^{ \rho \nu})$ \\ \hline 
\end{tabular}
\endgroup
\caption{List of relevant flavor universal dimension six operators.}
\label{tab:UniversalOps}
\end{table}

We first consider the operators in~\cref{tab:UniversalOps} that can be generated at tree level in minimally-coupled strong sectors. One can additionally consider the universal operators $\mathcal{O}_{2G}$ (tree level) and $\cO_{HW}, \cO_{HB}, \cO_{g}, \cO_{\gamma}$, $\mathcal{O}_{3W}$, $\mathcal{O}_{3G}$ (1-loop) as defined in~\cite{Giudice:2007fh,Contino:2013kra}. As we show in \cref{app:B}, these always give sub-leading constraints from EW precision data and we will not consider them further here. According to the rules of~\cref{eq:scalingEFT}, and defining $f = m_*/g_*$, the matching conditions for the universal operators are
\begin{align}
\cC_H &= \frac{c_H}{f^2}, & \cC_T &= \frac{c_T}{f^2}, & \cC_W &= \frac{c_W}{m_*^2}, & \cC_B &= \frac{c_B}{m_*^2}, & \cC_{2W} &= \frac{c_{2W}}{g_{*}^2 m_*^2}, & \cC_{2B} &= \frac{c_{2B}}{g_{*}^2 m_*^2},
\label{eq:Umatching}
\end{align}
where the small $c_{H,T,W,\dots}$ are model-dependent, $O(1)$ dimensionless parameters.

We will now briefly describe the phenomenological impact of these universal operators.
The operator $\cO_{H}$ modifies the Higgs kinetic term and thus universally rescales the coupling of the Higgs to all other fields, \eg $HVV$ couplings. The operator $\cO_T$ violates custodial symmetry, giving a tree-level contribution to the Peskin–Takeuchi parameter $\hat T = \alpha T = c_T v^2/f^2$ that imposes a very severe constraint on the scale of the model unless the strong sector respects custodial symmetry. The operators $\cO_{W,B}$ contribute to the $\hat S = \alpha S/(4 s_W^2)$ parameter as
\begin{equation}
\hat{S} = (c_W + c_B) \frac{m_W^2}{m_*^2}\,,
\end{equation}
which depends only on the overall scale $m_*$. One can understand this simple ratio of scales as mixing between the EW gauge bosons and their composite partners.
Finally, the operators $\cO_{2W}$ and $\cO_{2B}$ contribute to the oblique $W$ and $Y$ parameters, respectively
\begin{align}
W = c_{2W}\frac{g_2^2}{g_*^2}\frac{m_W^2}{m_*^2}\,, \hspace{17.5mm} Y = c_{2B}\frac{g_1^2}{g_*^2}\frac{m_W^2}{m_*^2}\,.
\end{align}
For the reader accustomed to working in the Warsaw basis, we see using equations of motion that $\cC_{W,B}$ are giving a flavor-universal shifts to operators of the form $( i H^\dagger  \overleftrightarrow {D^\mu} H)(\bar \psi \gamma_\mu \psi)$, namely flavor-universal corrections to the $W$ and $Z$ couplings to SM fermions. This is also the case for $\cC_{2W,2B}$, which additionally generate universal 4-fermion operators.

The oblique parameters $\hat S$, $\hat T$, $W$, $Y$ are all constrained by LEP data at the per-mille level~\cite{Barbieri:2004qk}. In composite Higgs models, however, we see that they come with a hierarchical power counting that yields $\hat T > \hat S > W,Y$ for $g_* > 1$. Thus, $\hat S$, $\hat T$, and Higgs coupling modifications from $\cC_H$ typically give the dominant universal constraints, unless $g_*$ is very small. For large $g_*$, we also see that loop contributions to $\hat T \propto (4\pi f)^{-2} $ can be of a similar size to tree-level contributions to $\hat S \propto (g_* f)^{-2}$. This basic observation is the fundamental reason behind the importance of the 1-loop effects we will consider in this paper. 

To conclude this section, we note that at the level of universal operators, we have only 2 free parameters ($g_*, m_*$) if we assume all unknown coefficients to be $O(1)$.

\subsection{Flavor non-universal operators from a strongly-interacting top quark}
\label{sec:SITQops}
\begin{table}[t]
\centering
\begingroup
\renewcommand*{\arraystretch}{1.4}
\begin{tabular}{|c|c|}
 \hline
  \multicolumn{2}{|c|}{\textbf{Flavor non-universal operators}} \\ \hline \hline 
 \multicolumn{2}{|c|}{EW vertex corrections} \\ \hline 
 $\mathcal{O}_{Hq}^{(1)}= ( H^{\dagger} i \overleftrightarrow{D}_{\mu} H ) (\bar{q}^3_L \gamma^{\mu} q^3_L)$ & $\mathcal{O}_{Hq}^{(3)}= ( H^{\dagger} i \overleftrightarrow{D}_{\mu}^I H )(\bar{q}^3_L \gamma^{\mu} \tau^I q^3_L)$ \\ \hline
 $\mathcal{O}_{Ht}= ( H^{\dagger} i \overleftrightarrow{D}_{\mu} H) (\bar{t}_R \gamma^{\mu} t_R)$ & $ \mathcal{O}_{tD} = g_1 (\bar{t}_R \gamma^{\mu} t_R) \partial^{\nu} B_{\mu \nu}$ \\ \hline
 $ \mathcal{O}_{qD}^{(1)}= g_1 (\bar{q}_L^3 \gamma^{\mu} q_L^3) \partial^{\nu} B_{\mu \nu}$ & $\mathcal{O}_{qD}^{(3)} = g_2 (\bar{q}_L^3 \gamma^{\mu} \tau^I q_L^3) D^{\nu} W^I_{\mu \nu}$ \\ 
 \hline \hline
 \multicolumn{2}{|c|}{4-fermion operators } \\ \hline 
 $\mathcal{O}_{qq}^{(1)} = (\bar q_L^3 \gamma^\mu q_L^3)(\bar q_L^3 \gamma_\mu q_L^3)$ & $ \mathcal{O}_{qq}^{(3)} = (\bar q_L^3 \gamma^\mu \tau^I q_L^3)(\bar q_L^3 \gamma_\mu \tau^I q_L^3)$ \\ \hline
 $\mathcal{O}^{(1)}_{qt} = (\bar q_L^3 \gamma^\mu q_L^3)(\bar t_R \gamma_\mu t_R)$ & $\mathcal{O}^{(8)}_{qt} = (\bar q_L^3 \gamma^\mu T^A q_L^3)(\bar t_R \gamma_\mu T^A t_R)$ \\ \hline
  \multicolumn{2}{|c|}{  $\mathcal{O}_{tt} = (\bar t_R \gamma^\mu t_R)(\bar t_R \gamma_\mu t_R)$ } \\ 
 \hline \hline
  \multicolumn{2}{|c|}{Dipoles and Yukawas} \\ \hline 
 $\mathcal{O}_{tB} = g_1 (\bar q^3_L \sigma^{\mu\nu} t_R) \widetilde{H} B_{\mu\nu}$ & $\mathcal{O}_{tW} = g_2 (\bar q^3_L \sigma^{\mu\nu} \tau^I t_R) \widetilde{H} W^I_{\mu\nu}$  \\ \hline
$\mathcal{O}_{tG} = g_3 (\bar q^3_L \sigma^{\mu\nu} T^A t_R) \widetilde{H} G^A_{\mu\nu}$ & $\mathcal{O}_{tH} =(H^\dagger H)(\bar q_L^3 \widetilde{H} t_R) $ \\ \hline \hline
\end{tabular}
\endgroup
\caption{List of relevant flavor non-universal dimension-six operators built from $q_L^3$ and $t_R$. In the absence of explicit flavor indices, $q = q^3$ is always implied. }
\label{tab:nonUniversalOps}
\end{table}
The leading deformation to the universal picture comes from
considering a strongly-interacting top quark, namely writing operators involving the fields $q_L^3$, $t_R$, and $H$. We have classified the relevant dimension-six interactions in~\cref{tab:nonUniversalOps}. Our notation is such that in the absence of explicit flavor indices, operators should always be understood to involve only purely third-family quarks, \ie $\cC_{Hq}^{(1)} = [\cC_{Hq}^{(1)}]_{33}$ and $\cC_{qq}^{(1)} = [\cC_{qq}^{(1)}]_{3333}$. According to \cref{eq:Lmix}, all non-universal effects are controlled by $\lambda_L$ and $\lambda_R$. It will be useful to define
\begin{align}
\eps_L = \frac{\lambda_L}{g_*}\,, \hspace{17.5mm} \eps_R = \frac{\lambda_R}{g_*}\,,
\end{align}
such that the top Yukawa is $y_t \simeq g_* \eps_L \eps_R$ according to~\cref{eq:YukEFT}. These parameters have the range $y_{t}/g_* \leq \eps_{L,R} \leq 1$. Because of the top Yukawa relation, the SILH+TQ scenario introduces only one additional parameter associated with the top quark mixing, \eg we can take ($g_*, m_*$,\,$\eps_R$).
Applying the rules of~\cref{eq:scalingEFT}, the matching conditions are
\begin{align}
\cC_{Hq}^{(1)} &= c_{Hq}^{(1)}\frac{\eps_L^2}{f^2}, & \cC_{Hq}^{(3)} &= c_{Hq}^{(3)}\frac{\eps_L^2}{f^2}, & \cC_{Ht} &= c_{Ht} \frac{\eps_R^2}{f^2}, & \cC_{tD} &= c_{tD}\frac{\eps_R^2}{m_*^2}, & \cC_{qD}^{(1)} &= c_{qD}^{(1)}\frac{\eps_L^2}{m_*^2}, \nn
\cC_{qD}^{(3)} &= c_{qD}^{(3)}\frac{\eps_L^2}{m_*^2}, & \cC_{qq}^{(1)} &= c_{qq}^{(1)}\frac{\eps_L^4}{f^2}, & \cC_{qq}^{(3)} &= c_{qq}^{(3)}\frac{\eps_L^4}{f^2}, & \cC_{qt}^{(1)} &= c_{qt}^{(1)}\frac{y_t^2}{m_*^2}, & \cC_{qt}^{(8)} &= c_{qt}^{(8)}\frac{y_t^2}{m_*^2}, \nn 
\cC_{tt} &= c_{tt}\frac{\eps_R^4}{f^2}, & \cC_{tB} &= \frac{g_*^2}{16\pi^2}\frac{ y_t c_{tB}  }{m_*^2}, & \cC_{tW} &= \frac{g_*^2}{16\pi^2}\frac{ y_t c_{tW} }{m_*^2}, & \cC_{tG} &= \frac{g_*^2}{16\pi^2}\frac{ y_t c_{tG}}{m_*^2}, & \cC_{tH} &= c_{tH}\frac{y_t}{f^2} \,,
\label{eq:NUmatching}
\end{align}
where we have used the top Yukawa relation $y_t \simeq g_* \eps_L \eps_R$ whenever the product of left- and right-handed fermions occurs in the operator. 

If unsuppressed, the operators in~\cref{eq:NUmatching} proportional to $1/f^2 = g_*^2/m_*^2$, would be expected to dominate the low-energy phenomenology. These are the EW vertex corrections $\cO_{Hq}^{(1,3)}$ and $\cO_{Ht}$, as well as 4-quark operators.\footnote{Also $\cO_{tH}$, but it is poorly constrained phenomenologically, see~\cref{tab:boundTable}.} However, tree-level EW vertex corrections can be protected by invoking $P_{LR}$, while 4-quark operators do not enter EW precision observables at tree level. Thus, these operators dominantly contribute via 1-loop RGE, with the leading effects due to top-quark loops behaving as $y_t^2 N_c \log(\Lambda^2/\EW^2)/(4\pi f)^2$. This puts non-universal contributions scaling as $1/f^2$ on the same footing as operators behaving as $1/m_*^2$, even if $g_*$ is not very large. Operators scaling as $1/m_*^2$ include the $S$-parameter, as well the operators $\cO_{qD}^{(1,3)}$ and $\cO_{tD}$ that involve the product of a third-family quark current with a universal one represented by $D^\nu F_{\mu\nu}^{\rm SM}$. For the dipoles, we have assumed a strong dynamics loop suppression $(g_*/4\pi)^2$. If this is not the case, we have checked in~\cref{app:B} that the dipoles give a similar bound to the $S$-parameter.\footnote{The leading constraint comes from $\cO_{tW}$, where high-$p_T$ searches give a bound around 3 TeV~\cite{Ethier:2021bye}. This operator is therefore an interesting target for the HL-LHC to probe models with unsuppressed dipoles.} Since they do not introduce any new bound or parameter scaling, we neglect them going forward. 

The three options we consider for the SITQ mixing are
\begin{enumerate}[i)]
\centering
    \item \textbf{Left compositeness:} $\eps_L \rightarrow 1$, $\eps_R \rightarrow y_t/g_*$ \,,
    \item \textbf{Mixed compositeness:} $\eps_L = \eps_R \rightarrow \sqrt{y_t/g_*}$ \,,
    \item \textbf{Right compositeness:} $\eps_L \rightarrow y_t/g_*$, $\eps_R \rightarrow 1$ \,,
\end{enumerate}
which as we will see in~\S~\ref{sect:CHmodels}, will give us a comprehensive picture of how all bounds in $(m_*, g_*)$ plane depend on the SITQ mixing choice.

\subsubsection*{Top partners}
Before concluding this section, let us comment on a possibility that is frequently considered in the literature, which is that of light top partners. This possibility is attractive for obtaining a light Higgs mass, which receives a finite loop contribution behaving as $m_H^2 \propto N_c y_T^2 m_T^2 / (4\pi)^2$. In principle, top partners with mass $m_T \ll m_*$ would violate our one coupling, one scale hypothesis. However, since LHC bounds require $m_T > 1$ TeV, we expect that our EFT description still applies. According to Ref.~\cite{deBlas:2017xtg}, top partners that couple linearly to the SM can only generate EW vertex modifications $\cO_{Hq}^{(1,3)}$, $\cO_{Ht}$ and top Yukawa corrections $\cO_{tH}$ at tree level, with expected WCs of $y_T^2/m_T^2$. Therefore, top partners do not introduce any new operators in our EFT, and would simply change the coefficients in~\cref{eq:NUmatching} by $\lesssim O(1)$ as long as $m_T^2  \gtrsim y_T^2 f^2$. This seems to be quite a reasonable condition if $y_T$ is not large, which would anyway defeat the purpose of light top partners controlling the contribution to $m_H^2$. For more discussion, see \eg~\cite{Pomarol:2008bh,Matsedonskyi:2012ym,Grojean:2013qca,Panico:2015jxa}.

\subsection{Leading flavor-violating operators}
\label{sec:FVops}
Next, we discuss the most important flavor-violating operators. Inserting the $\lambda_q^i \equiv g_* \eps_q^i$ spurion via the rules of~\cref{eq:scalingEFT}, one expects $\Delta F = 1,2$ operators of the form
\begin{align}
\mathcal{L}_{\rm EFT} &\supset \frac{(\eps_q^i \eps_L^*)}{f^2} ( H^{\dagger} i \overleftrightarrow{D}_{\mu} H ) (\bar{q}^i_L \gamma^{\mu} q^3_L) + \frac{(\eps_q^i \eps_L^*)^2}{f^2} (\bar q_L^i \gamma_\mu q_L^3)^2 + \frac{g_*^2}{16\pi^2}\frac{\eps_q^i}{\eps_L} \frac{y_b \, g_{\rm SM}}{m_*^2} (\bar q_L^i \sigma^{\mu\nu} H b_R) F^{\rm SM}_{\mu\nu} \,,
\label{eq:LagFV}
\end{align}
in the low-energy EFT. To obtain the correct CKM mixing we should have $\eps_q^i \simeq \eps_L V_{ti}^{*}$. The dipole operator gives rise to $b\rightarrow s\gamma$ transitions. Though it is $\propto y_b$, we have included it since this is one of the leading constraints for $y_b \neq 0$.\footnote{We assume a partial compositeness origin for $y_b \simeq g_* \eps_L \eps_b $ to exchange $\eps_b$ for $y_b$. } We have also assumed suppression by a loop in the strong dynamics.\footnote{In principle, dipoles could be generated at tree-level by pseudovector states~\cite{Cata:2014fna}. If this is case, then $b\rightarrow s\gamma$ transitions would give a constraint of $m_* \gtrsim 6$ TeV, see also~\cite{Glioti:2024hye}.} With $y_b$ and this loop factor, $b\rightarrow s\gamma$ transitions give a bound $f \gtrsim  0.5$ TeV~\cite{Davighi:2023evx,Allwicher:2023shc}. In~\S~\ref{sect:CHmodels}, we will find a lower bound of $f \gtrsim 1$ TeV, so these effects are under control and we do not consider them further here. We now analyze the leading phenomenological impact of the remaining two operators.

\subsubsection*{$B_s \rightarrow \mu^+ \mu^-$ decays}
A very important contribution to $B_s \rightarrow \mu^+ \mu^-$ comes from flavor-violating $\cC_{Hq}$ operators such as the first term in~\cref{eq:LagFV} that induce effective $bs$ vertices for the SM $Z$. One can understand this as the $Z$ boson acquiring the flavor-violating couplings of its heavy composite partner via mixing. The result is a nearly purely axial vector contribution $C_{10}$ to the $bsll$ system, so it is useful to define
\begin{equation}
C_{10}^U = \frac{\pi v^2}{\alpha} \frac{1}{V_{tb} V_{ts}^*} \left([\cC_{Hq}^{(1)}]_{23} + [\cC_{Hq}^{(3)}]_{23}\right) \,,
\end{equation}
where the superscript $U$ indicates a universal contribution to all leptons. We estimate
\begin{equation}
C_{10}^U = \frac{\pi v^2}{\alpha} \frac{1}{V_{tb} V_{ts}^*} \frac{\eps_q^i \eps_L}{f^2} \left([c_{Hq}^{(1)}]_{23} + [c_{Hq}^{(3)}]_{23}\right) \simeq \frac{\pi v^2}{\alpha} \frac{\eps_L^2}{f^2}\,,
\label{eq:C10U}
\end{equation}
where $\alpha = e^2 / 4\pi$. We get the bound
\begin{equation}
m_* \gtrsim \frac{y_t v}{\eps_R} \sqrt{\frac{\pi}{\alpha C_{10}^U }} \,.
\end{equation}
The analysis in~\cite{Greljo:2022jac} gives $C_{10}^U =0.21 \pm 0.19$ from a single operator fit to the $B_s \rightarrow \mu^+ \mu^-$ data. In the best case scenario of a positive contribution, we find
\begin{equation}
m_* \gtrsim 6.7~\text{TeV} \frac{y_t}{\eps_R} \,,
\end{equation}
in agreement with the lower bound quoted in~\cite{Glioti:2024hye} based on the analysis in~\cite{Ciuchini:2022wbq}. Using $y_t(\Lambda = 2.5~\text{TeV}) = 0.82$, we get $m_* \gtrsim 5.5~\text{TeV}/\eps_R$. Note that the stringent bound on $B_s\rightarrow \mu^+ \mu^-$ forbids any large enhancement in $b\rightarrow s\nu\bar\nu$ transitions~\cite{Buras:2014fpa,Crosas:2022quq,Lizana:2023kei,Davighi:2023evx}. We note that this bound actually has the same parametric scaling as $Z b\bar b$ vertex modifications, since they are both in the same $\cO_{Hq}$ class of operators with scaling $\eps_L^2 /f^2$. This also means that the same $P_{LR}$ symmetry that protects $Z b\bar b$ would enforce $[c_{Hq}^{(1)}]_{23} = -[c_{Hq}^{(3)}]_{23}$ in~\cref{eq:C10U}, also protecting tree-level contributions to $B_s \rightarrow \mu^+ \mu^-$~\cite{Barbieri:2019kfa}.

\subsubsection*{$B$-meson mixing}
Next is $B$-meson mixing, induced by the second operator in~\cref{eq:LagFV}. The NP contribution normalized to the SM one can be written as
\begin{equation}
|\Delta C_{B_{i}}| = \bigg|\frac{C_{B_{i}}^{\rm NP} }{C_{B_{i}}^{\rm SM} }\bigg|= \frac{4\pi^2}{G_F^2 m_W^2 S_0(x_t)} \bigg|\frac{[\cC_{qq}^{(1)}]_{i3i3} + [\cC_{qq}^{(3)}]_{i3i3}}{(V_{tb} V_{ti}^*)^2}\bigg| \simeq \frac{4\pi^2}{G_F^2 m_W^2 S_0(x_t)}\frac{\eps_L^4}{f^2} \,,
\end{equation}
where $S_0(x_t) = 2.37$~\cite{Buchalla:1995vs}. Using the latest results in~\cite{Bona:2022wa}, we find $|\Delta C_{B_{d}}| < 0.27$ and $|\Delta C_{B_{s}}| < 0.19$ at 95\% confidence level, so the strongest bound comes from the $B_s$ system. In terms of a bound on $m_*$, we find
\begin{equation}
m_* \gtrsim 10~\text{TeV}\frac{y_t^2}{g_* \eps_R^2} \,,
\end{equation}
and again taking $y_t(\Lambda = 2.5~\text{TeV}) = 0.82$, we get $m_* \gtrsim 6.7~\text{TeV}/(g_* \eps_R^2)$.

%----------------------------------------------------------------------------
\section{Bounds on the effective operators from EW precision data}
\label{sect:singleOps}
\begin{table}[t]
\centering
\begin{subtable}{0.45\textwidth}
    \centering
\begin{tabular}{|c|c|c|}
\hline
Wilson Coef. & [Obs]$_{\text{bound}}$ & $\Lambda_{\text{bound}}$ [TeV] \\
\hline
$\mathcal{C}_T$ & $A_b^{\text{FB}}$ & 8.17 \\ \hline
$\mathcal{C}_{Hq}^{(1)}$ & $R_{\tau}$ & 3.98 \\ \hline
$\mathcal{C}_{Hq}^{(3)}$ & $R_b$ & 3.94 \\ \hline
$\mathcal{C}_{Ht}$ & $A_b^{\text{FB}}$ & 3.00 \\ \hline
$\cC_{Hq}^{(-)}$ & $A_b^{\text{FB}}$ & 2.98 \\ \hline
$\mathcal{C}_B$ & $A_b^{\text{FB}}$ & 2.48 \\ \hline
$\mathcal{C}_W$ & $A_b^{\text{FB}}$ & 2.41 \\ \hline
$\mathcal{C}_{tW}$ & $A_b^{\text{FB}}$ & 1.86 \\ \hline
$\mathcal{C}_{qD}^{(3)}$ & $R_{\tau}$ & 1.83 \\ \hline
$\mathcal{C}_{qq}^{(1)}$ & $R_{\tau}$ & 1.50 \\ \hline
$\mathcal{C}_{tB}$ & $A_b^{\text{FB}}$ & 1.44 \\ \hline
$\mathcal{C}_{2W}$ & $A_b^{\text{FB}}$ & 1.29 \\ \hline
$\mathcal{C}_{qt}^{(1)}$ & $R_{\tau}$ & 1.14 \\ \hline
$\mathcal{C}_{qD}^{(1)}$ & $A_b^{\text{FB}}$ & 1.12 \\ \hline
$\mathcal{C}_{tt}$ & $A_b^{\text{FB}}$ & 1.05 \\ \hline
$\mathcal{C}_{qq}^{(3)}$ & $R_b$ & 0.94 \\ \hline
$\mathcal{C}_{tD}$ & $A_b^{\text{FB}}$ & 0.93 \\ \hline
$\mathcal{C}_{2B}$ & $A_b^{\text{FB}}$ & 0.77 \\ \hline
$\mathcal{C}_H$ & $A_b^{\text{FB}}$ & 0.47 \\ \hline
$\mathcal{C}_{tG}$ & $A_b^{\text{FB}}$ & 0.46 \\ \hline
$\mathcal{C}_{tH}$ & $H \to \mu \mu$ & 0.18 \\ \hline
$\mathcal{C}_{qt}^{(8)}$ & $R_{\tau}$ & 0.11 \\ \hline
\end{tabular}
    \caption{Current bounds}
    \label{tab:currentBounds}
\end{subtable}\hfill
\begin{subtable}{0.45\textwidth}
    \centering
\begin{tabular}{|c|c|c|}
\hline
Wilson Coef. & [Obs]$_{\text{bound}}$ & $\Lambda_{\text{bound}}$ [TeV] \\
\hline
$\mathcal{C}_T$ & $m_W$ & 74.24 \\ \hline
$\mathcal{C}_{Hq}^{(1)}$ & $m_W$ & 39.82 \\ \hline
$\mathcal{C}_{Hq}^{(3)}$ & $R_{\mu}$ & 24.81 \\ \hline
$\mathcal{C}_{Ht}$ & $m_W$ & 35.92 \\ \hline
$\cC_{Hq}^{(-)}$ & $m_W$ & 33.97 \\ \hline
$\mathcal{C}_B$ & $A_e$ & 26.15 \\ \hline
$\mathcal{C}_W$ & $A_e$ & 24.67 \\ \hline
$\mathcal{C}_{tW}$ & $A_e$ & 26.19 \\ \hline
$\mathcal{C}_{qD}^{(3)}$ & $R_{\mu}$ & 11.73 \\ \hline
$\mathcal{C}_{qq}^{(1)}$ & $m_W$ & 16.59 \\ \hline
$\mathcal{C}_{tB}$ & $A_e$ & 20.24 \\ \hline
$\mathcal{C}_{2W}$ & $A_e$ & 12.48 \\ \hline
$\mathcal{C}_{qt}^{(1)}$ & $m_W$ & 14.61 \\ \hline
$\mathcal{C}_{qD}^{(1)}$ & $A_e$ & 13.73 \\ \hline
$\mathcal{C}_{tt}$ & $m_W$ & 14.64 \\ \hline
$\mathcal{C}_{qq}^{(3)}$ & $R_{\mu}$ & 7.95 \\ \hline
$\mathcal{C}_{tD}$ & $A_e$ & 12.90 \\ \hline
$\mathcal{C}_{2B}$ & $A_e$ & 8.58 \\ \hline
$\mathcal{C}_H$ & $m_W$ & 6.03 \\ \hline
$\mathcal{C}_{tG}$ & $A_e$ & 7.91 \\ \hline
$\mathcal{C}_{tH}$ & \text{$H \to \tau \tau$} & 0.95 \\ \hline
$\mathcal{C}_{qt}^{(8)}$ & $m_W$ & 1.61 \\ \hline
\end{tabular}
    \caption{FCC-ee projection}
    \label{tab:FccBounds}
\end{subtable}
\caption{Current (left) and projected future (right) 95\% CL single operator bounds from EW precision data on the high-scale Wilson coefficients $\cC_i(\Lambda)$ of the operators defined in~\cref{tab:UniversalOps,tab:nonUniversalOps}. We also show the bound on $\cC_{Hq}^{(-)}$, defined as the WC of the operator $\cO_{Hq}^{(-)} = \cO_{Hq}^{(1)}-\cO_{Hq}^{(3)}$. For current (future) data we take $\Lambda = 2.5$ TeV (25 TeV). }
\label{tab:boundTable}
\end{table}
We have constructed our own EW likelihood which does not make any flavor assumptions. This likelihood has been previously applied and validated in~\cite{Allwicher:2023aql,Davighi:2023evx,Allwicher:2023shc}. We include all the $Z$- and $W$-pole observables listed in Tables 1
and 2 of~\cite{Breso-Pla:2021qoe}, using the same SM predictions\footnote{The only difference is that we include a 4 MeV theory error on $m_W$~\cite{Awramik:2003rn}.}, experimental values, and the same $G_F$, $m_Z$, and $\alpha$ input scheme. We do not include the latest CDF II result on $m_W$ \cite{CDF:2022hxs}. We also include Higgs decay data from~\cite{ParticleDataGroup:2022pth}. We calculated NP contributions to these observables in terms of a shift of the $W$ mass, $\delta m_W$, as well as shifts of the $W$ and $Z$ couplings to SM fermions described by $3 \times 3$ matrices $\delta g^{W \ell}$, $\delta g^{W q}$, $\delta g^{Z \, f}_L$, and $\delta g^{Z f}_R$ in Appendix C.1 of Ref.~\cite{Allwicher:2023aql}. We also provide tree-level expressions for $\delta m_W$ and $\delta g^{W,Z}$ in terms of SMEFT WCs in the Warsaw basis, which allows us to express the NP contribution to all pole observables in terms of SMEFT WCs. We then fit directly to the $Z$- and $W$-pole observables via a simple $\chi^2$ analysis, as described in~\cite{Allwicher:2023aql,Allwicher:2023shc}.

\subsection{Running the EW likelihood}
At the EW scale, only 23 of all the 2499 dimension-six SMEFT operators contribute to $Z$ and $W$-pole observables. However, as we showed in~\cite{Allwicher:2023shc} in the case of a $U(2)^5$-symmetric SMEFT~\cite{Faroughy:2020ina}, 120 out of 124 possible $U(2)^5$-invariant WCs can mix via RGE into the $Z$ and $W$-pole observables. While the $U(2)^5$ assumption may seem restrictive, it is exactly applicable here as all our effects are either flavor universal, or non-universal in the third family only. We account for RGE by running up our entire EW likelihood from $\EW = m_t$ to a high scale $\Lambda$, the scale at which we assume the NP is integrated out and matching to the SMEFT occurs. The choice for $\EW$ is discussed in~\S~\ref{sec:NLL}. We are able to do this analytically in the WCs using the evolution matrix method (EMM) implemented in \texttt{DSixTools} \cite{Celis:2017hod,Fuentes-Martin:2020zaz}. It is very important to emphasize that the EMM provides an excellent approximation to integrating the full set of coupled RG equations, so we are able to work analytically while simultaneously resumming any large logarithms and fully capturing operator mixing occurring beyond the first LL approximation. In this work we use two benchmark points for the high scale $\Lambda$, based on whether we are analyzing current data or doing a future projection for FCC-ee:
\begin{equation}
\Lambda = 
 \begin{cases} 
      2.5~\text{TeV} & \text{(current data)} \\
      25~\text{TeV} & \text{(FCC-ee projection)}
   \end{cases} \,
\end{equation}
The current choice is motivated by LHC bounds on the mass of top partners and other resonances, while both are motivated by the lower bound we obtain for $m_*$ in the two cases.

\subsection{Details of the future projection for FCC-ee/Tera-$Z$}
A tera-$Z$ machine would provide $O(10^5)$ more $Z$-bosons than LEP, so the naive statistical improvement on the uncertainty of EW precision observables could be up to a factor of $\approx 300$. Due to systematic and theoretical uncertainties, however, in practice a factor of 10-100 (10) is typically expected for leptonic (hadronic) observables.
A future projection for a tera-$Z$ circular $e^+ e^-$ collider was compiled by us in~\cite{Allwicher:2023shc}, which we will use again here. We report it in~\cref{sec:appFCC} for convenience. Our focus is on the FCC-ee program, but our projections should be generally applicable to other tera-$Z$ machines, such as the proposed CEPC project. We construct our projected EW likelihood for FCC-ee assuming a null result, \ie we set the experimental value of each observable to its corresponding SM theory prediction\footnote{We take the SM theory predictions from~\cite{Breso-Pla:2021qoe}.} and assume the relative error is reduced as in the ``Proj. Error Reduction" column of~\cref{tab:FCCeePROJZpole,tab:FCCeePROJother}. We neglect theoretical uncertainties for our main results, assuming that a vigorous theory effort will ultimately be able to match experimental precision (``TH3" scenario of~\cite{Blondel:2018mad}).  We note that with the information given here, our projection is fully reproducible. Since we will find the observables $A_e$ and $m_W$ to be the most constraining, the only difference with respect to Ref.~\cite{Allwicher:2023shc} is that we use the slightly more conservative projection for $A_e$ taken from~\cite{deBlas:2022ofj} (both have the same $m_W$ projection). For completeness, we also show in~\cref{app:C} how our results would be impacted if we assume some intrinsic theoretical uncertainty on these measurements.

\subsection{EFT operator bounds and future projections}
\label{sec:EFTopBounds}
Having discussed the EW likelihood, we are now ready to apply it in order to obtain bounds on the SILH+TQ operators. In~\cref{tab:boundTable}, we show the bounds we get by switching on each operator in~\cref{tab:UniversalOps,tab:nonUniversalOps} one at a time. We show the 95\% CL bounds both from current data, as well as the FCC-ee projection, sorted by the strength of the current bound (we keep this ordering in~\cref{tab:FccBounds} for ease of comparison). In each case, we also show the observable $[\text{Obs}]_{\rm bound}$ with the largest pull. As is well known, the current EW data has some mild $\approx 2\sigma$ tensions (see \eg~\cite{deBlas:2021wap}), resulting in some asymmetric intervals and even a mild preference for NP in some cases. To be conservative, we always choose the sign of the WC that gives the weakest bound, but for completeness we also report the bound in both directions in~\cref{app:B}. It is easy to verify there that the lower bound displayed in~\cref{tab:boundTable} never corresponds to a preference for NP. For ease of interpretation, we report the bound as an effective scale such that the WC should satisfy $|\cC| \leq 1/\Lambda_{\rm bound}^2$.

We understand that single operator bounds are qualitative in the sense that full models always give correlations between various WCs. Since we do not want to commit here to any particular model, we anyway think these single operator bounds are extremely useful to set the stage, as one can easily re-derive some key results long known in the literature. For example, one can immediately see the operators to be avoided if one wants to build a low-scale model. These are the first four operators in~\cref{tab:currentBounds}: $\cO_T$, $\cO_{Hq}^{(1)}$, $\cO_{Hq}^{(3)}$, and $\cO_{Ht}$. The first operator $\cO_T$ is absent for a custodially-symmetric strong sector. The next two non-universal operators result in tree-level shifts of the vertex $Z\rightarrow b_L \bar b_L \propto \cC_{Hq}^{(1)} + \cC_{Hq}^{(3)}$, giving a bound of around 4 TeV. The final operator $\cO_{Ht}$, which modifies the $Z$ coupling to $t_R$, does not give any strong bound at tree level, but as we will see in~\S~\ref{sec:LLrunningHq}, it has a large first LL running into the $T$ parameter. Even though this is a 1-loop effect, it still gives a strong bound of 3 TeV due to the per-mille precision at which $T$ is measured. 

Here it is important to note that invoking the $P_{LR}$ symmetry can simultaneously protect the $Z b_L \bar b_L$ and $Z t_R \bar t_R$ vertices, corresponding to $\cC_{Hq}^{(3)} = -\cC_{Hq}^{(1)}$ and $\cC_{Ht} = 0$ at tree level in the SMEFT. Thus, a custodially-symmetric strong sector together with fermion embeddings that realize $P_{LR}$ avoids the worst four bounds in~\cref{tab:boundTable}. For further analysis, it is useful to define the linear combinations
\begin{align}
\cO_{Hq}^{(+)} &= \cO_{Hq}^{(1)}+\cO_{Hq}^{(3)} \,, \\
\cO_{Hq}^{(-)} &= \cO_{Hq}^{(1)}-\cO_{Hq}^{(3)} \,,
\end{align}
that correct the vertices of down- and up-type quarks, respectively. Invoking $P_{LR}$ causes the WC of $\cO_{Hq}^{(+)}$ (that we denote as $\cC_{Hq}^{(+)}$) to vanish at tree level, while in general we still have $\cC_{Hq}^{(-)} \neq 0$. This corresponds to unprotected corrections to the $Z t_L \bar t_L$ vertex, which also run at first LL into the $T$ parameter. Indeed, one sees that the bound on $\cC_{Hq}^{(-)}$ is similar to the one on $\cC_{Ht}$.\footnote{In fact at leading log they have identical bounds- the small difference comes from resummation.} 

This is our first glimpse into why non-universal RGE effects are very important probes of these symmetry-protected models. The bound on $\cC_{Hq}^{(-)}$ coming from RGE is comparable to the leading tree-level universal bound from $S \propto \cC_{B} + \cC_W$. Below those,~\cref{tab:boundTable} shows several flavor non-universal operators with bounds around 1-2 TeV, all of which come from RG mixing into EW precision observables. At FCC-ee, the projected overall improvement is about one order of magnitude, translating into several operators with naive bounds of around 10-20 TeV. Let us now discuss the most significant of these RG effects in detail.

\section{Discussion of RGE effects due to the non-universal operators}
\label{sec:RGEeffects}
Many operators shown in~\cref{tab:boundTable}, such as 4-quark operators, do not affect EW precision observables at tree level. They are constrained  nonetheless because they mix into operators that are strongly constrained on the $Z$ and $W$ poles as we perform the RGE from the high scale $\Lambda$ down to $\EW$. Overall, there are three important categories: i) first leading-log running in $y_t$ of 4-top operators $\cO_{4q}$ into EW vertex modifications $\cO_{Hq}$, ii) first leading-log running in $y_t$  of EW vertex modifications $\cO_{Hq}$ into the custodial-violating operator $\cO_{HD} = |H^\dagger D_\mu H|^2 \propto T$, and iii) 2-loop double-log mixing $\cO_{4q} \rightarrow \cO_{Hq} \rightarrow \cO_{HD}$ that can be thought of as the product of the two first LL processes in i)+ii). We now discuss each of these in turn, comparing the log-enhanced RGE with standard estimates for 1-loop finite matching contributions.
\subsection{First leading log running of third-family 4-quark operators}
\label{sec:LLrunningqq}
\begin{figure}[t]
    \centering
    \begin{tikzpicture}[thick,>=stealth,scale=1]
        \node[wc] at (0,0) {};
        \node[below=7pt] at (0,0) {$\cC_{4q}$};
        \draw[midarrow] (0,0) -- (-1,1) node[left] {$q$};
        \draw[midarrow] (-1,-1) node[left] {$q$} -- (0,0);
        \draw[midarrow] (1,-1) -- (0,0);
        \draw[midarrow] (0,0) -- (1,1);
        \draw[midarrow] (1,1) -- (1,-1);
        \draw[dashed] (1,1) -- (2,1) node[right] {$H$};
        \draw[dashed] (1,-1) -- (2,-1) node[right] {$H$};
        \begin{scope}[xshift=7.5cm]
            \draw[midarrow] (-0.9,0) -- (-1.8,0.9) node[left] {$q$};
            \draw[midarrow] (-0.9,0) -- (-1.8,-0.9) node[left] {$q$};
            \draw[midarrow] (-0.9,0) arc (180:0:0.9);
            \draw[midarrow] (0.9,0) arc (0:-180:0.9);
            \draw[dashed] (0.9,0) -- (1.8,0.9) node[right] {$H$};
            \draw[dashed] (0.9,0) -- (1.8,-0.9) node[right] {$H$};
            \node[wc] at (0.9,0) {};
            \node[wc] at (-0.9,0) {};
        \end{scope}
    \end{tikzpicture}
    \caption{Left: Logarithmically divergent graph giving the RG mixing of $\cO_{4q} \rightarrow \cO_{Hq}$. Right: Quadratically divergent graph from insertions of $\cC_{4q}$ and $\cC_{Hq}$ in the EFT, corresponding to a model-dependent finite contribution to $\cC_{Hq}$ in the full UV theory. See~\S~\ref{sec:LLrunningqq} for details.
    \label{fig:4quarksLL}}
\end{figure}
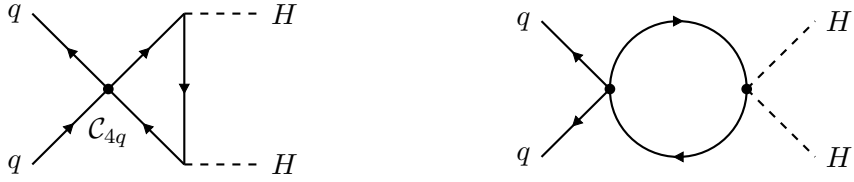
Via two insertions of $y_t$, the 4-top operators $\cO_{4q} := \cO_{qq}^{(1)},\cO_{qq}^{(3)}, \cO_{qt}^{(1)},\cO_{tt}$ run at first leading log into $\cO_{Hq}^{(1,3)}$ and $\cO_{Ht}$ via the diagrams shown on the left side of~\cref{fig:4quarksLL}. The divergent parts of these graphs are of course model independent and can be easily extracted from the 1-loop RG equations~\cite{Jenkins:2013wua}. We find the following
\begin{align}
[\cC_{Hq}^{(1)}]_{\rm LL} &= \frac{y_t^2}{16\pi^2} \left[(1+2N_c) \cC_{qq}^{(1)} + 3 \cC_{qq}^{(3)} -N_c \cC_{qt}^{(1)} \right] \log\left(\frac{\mu^2}{\Lambda^2}\right)\,, \\
[\cC_{Hq}^{(3)}]_{\rm LL} &= \frac{y_t^2}{16\pi^2} \left[(1-2N_c) \cC_{qq}^{(3)} - \cC_{qq}^{(1)} \right] \log\left(\frac{\mu^2}{\Lambda^2}\right)\,,\\
[\cC_{Ht}]_{\rm LL} &= \frac{y_t^2}{16\pi^2} \left[N_c \cC_{qt}^{(1)} - 2(1+N_c) \cC_{tt}\right] \log\left(\frac{\mu^2}{\Lambda^2}\right)\,,
\end{align}
where $N_c = 3$ is the number of QCD colors.
These log-enhanced terms should be compared against potential finite contributions from the UV, represented by the quadratically divergent graphs on the right side of ~\cref{fig:4quarksLL} (there are also graphs $\propto [\cC_{Hq}^{(1)}]^2$). For example, via NDA we estimate the following finite parts\footnote{Without $P_{LR}$, there would similarly be finite contributions $[\cC_{Ht}]_{\rm finite} \propto \lambda_R^4$ from insertions of $\cC_{Ht}$.}
\begin{align}
[\cC_{Hq}^{(1)}]_{\rm finite}  &\simeq -\frac{N_c\Lambda^2}{16\pi^2} [\cC_{Hq}^{(1)}][\cC_{qq}^{(1)}]  \approx -\frac{N_c\lambda_{L}^4 \eps_L^2}{16\pi^2 m_*^2} \,, \\ [\cC_{Hq}^{(1)}]_{\rm finite}  &\simeq \frac{\Lambda^2}{16\pi^2} [\cC_{Hq}^{(1)}]^2  \approx \frac{\lambda_{L}^4}{16\pi^2 m_*^2}  \,,
\label{eq:CHqFP}
\end{align}
where we have taken the naive cutoff $\Lambda = m_*$.
The most important thing to notice is that, according to~\cref{eq:NUmatching}, the RG contributions behave as $y_t^2 \lambda_L^2 \log(\mu^2/\Lambda^2) $ vs. $\lambda_L^4$ of the finite parts. For large $\lambda_L$, finite parts can therefore be important. However, other constraints, \eg from flavor, typically require that $\lambda_L$ is not too large. In addition, we will see next that finite contributions to $\cC_{HD}$ always dominate over $[\cC_{Hq}^{(1)}]_{\rm finite}$ due to the tighter bound on $T$.

\subsection{First leading log running of third-family EW vertex corrections}
\label{sec:LLrunningHq}
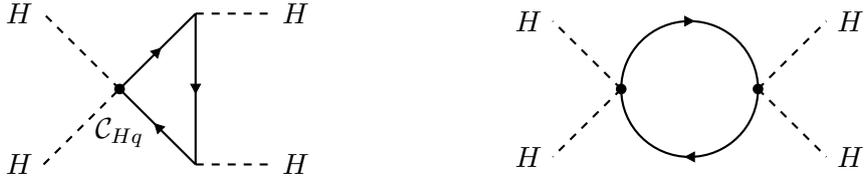
\begin{figure}[t]
    \centering
    \begin{tikzpicture}[thick,>=stealth,scale=1]
            \node[wc] at (0,0) {};
            \node[below=7pt] at (0,0) {$\cC_{Hq}$};
            \draw[dashed] (0,0) -- (-1,1) node[left] {$H$};
            \draw[dashed] (-1,-1) node[left] {$H$} -- (0,0);
            \draw[midarrow] (1,-1) -- (0,0);
            \draw[midarrow] (0,0) -- (1,1);
            \draw[midarrow] (1,1) -- (1,-1);
            \draw[dashed] (1,1) -- (2,1) node[right] {$H$};
            \draw[dashed] (1,-1) -- (2,-1) node[right] {$H$};
        \begin{scope}[xshift=7.5cm]
            \draw[dashed] (-0.9,0) -- (-1.8,0.9) node[left] {$H$};
            \draw[dashed] (-0.9,0) -- (-1.8,-0.9) node[left] {$H$};
            \draw[midarrow] (-0.9,0) arc (180:0:0.9);
            \draw[midarrow] (0.9,0) arc (0:-180:0.9);
            \draw[dashed] (0.9,0) -- (1.8,0.9) node[right] {$H$};
            \draw[dashed] (0.9,0) -- (1.8,-0.9) node[right] {$H$};
            \node[wc] at (0.9,0) {};
            \node[wc] at (-0.9,0) {};
        \end{scope}
    \end{tikzpicture}
    \caption{Left: Logarithmically divergent graph giving the RG mixing of $\cO_{Hq} \rightarrow \cO_{HD}$. Right: Quadratically divergent graph from a double insertion of $\cC_{Hq}$ in the EFT, corresponding to a model-dependent finite contribution to $\cC_{HD}$ in the full UV theory. See~\S~\ref{sec:LLrunningHq} for details.
    \label{fig:HqLL}}
\end{figure}

Due to the graphs on the left side of~\cref{fig:HqLL}, the operators $\cO_{Hq} := \cO_{Hq}^{(1)}, \cO_{Ht}$ run at first leading log via two insertions of the top Yukawa into $\cO_{HD}$, which gives the $T$ parameter in the Warsaw basis. That the $T$ parameter can be generated by the running is due to the fact that $y_t \gg y_b$ badly breaks custodial symmetry.\footnote{One can explicitly see the breaking of custodial symmetry due to the SM Yukawa couplings in~\cite{Jenkins:2013wua}. The contribution to $\cC_{HD}$ depends on the linear combination $Y_u Y_u^\dagger \cC_{Hu} - Y_d Y_d^\dagger \cC_{Hd} - [Y_u Y_u^\dagger - Y_d Y_d^\dagger]\cC_{Hq}^{(1)}$.} We find
\begin{equation}
[\cC_{HD}]_{\rm LL} = \frac{N_c y_t^2}{4\pi^2} \left[\cC_{Hq}^{(1)} - \cC_{Ht} \right] \log\left(\frac{\mu^2}{\Lambda^2}\right)\,,
\label{eq:CHDLL}
\end{equation}
with $\hat T = \alpha T =  - v^2 \cC_{HD}/2$ in the Warsaw basis.
For comparison, we again use NDA to estimate finite contributions to $\cC_{HD}$ from the quadratically divergent graph on the right side of~\cref{fig:HqLL}
\begin{align}
[\cC_{HD}]_{\rm finite} &\simeq -\frac{N_c \Lambda^2}{16\pi^2} [\cC_{Hq}^{(1)}]^2 \approx -\frac{N_c \lambda_{L}^4}{16\pi^2 m_*^2}  \,,
\label{eq:CHDfinite}
\end{align}
where we have taken $\Lambda = m_*$. We note that the naive cutoff analysis giving~\cref{eq:CHqFP,eq:CHDfinite} agrees with the typical estimates for the finite contributions found in the literature~\cite{Giudice:2007fh,Panico:2015jxa}. According to~\cref{eq:NUmatching}, we have $C_{Hq}^{(1)}\simeq \eps_L^2/f^2$, so this finite contribution is subleading to the log-enhanced RGE in the case of a composite $t_R$ ($\eps_R \rightarrow 1$). However, this is the dominant finite contribution, and it can be important in the mixed and left compositeness cases. We thus include the estimate for $\hat T_{\rm finite}$ coming from~\cref{eq:CHDfinite} in our result plots of~\S~\ref{sec:mainResults}, even though it is never dominant compared to other constraints.

\emph{It is important to emphasize that all finite contributions are quadratically sensitive to the UV, while the RGE is model independent.} Here, we have taken a conservative approach choosing a naive cutoff of $\Lambda = m_*$ to estimate the finite parts. Even so, we have shown that they remain under control compared to the RG contributions. In some models, for example with light top partners~\cite{Matsedonskyi:2012ym,Grojean:2013qca}, it can be possible to have $\Lambda < m_*$. This would quadratically suppress the finite parts, while the RG contributions would change only logarithmically. This shows that tree-level matching combined with 1-loop running is a sensible approach for assessing the leading 1-loop effects in composite Higgs scenarios in a model-independent manner.

\subsection{Two-loop double-log contributions of third-family 4-quark operators}
\label{sec:NLL}
We turn now to perhaps the most important point of this paper. There is an anomalously large 2-loop LL contribution $\cO_{4q} \rightarrow \cO_{Hq} \rightarrow \cO_{HD}$ coming from the product of the two graphs in~\cref{fig:NLLrunning}. If we neglect the running of the SM couplings as described in Appendix B of~\cite{Allwicher:2023aql}, we can estimate this double-log effect analytically
\begin{align}
[\cC_{HD}]_{{\rm log^2}}\approx
\frac{2 N_c \, y_t^4}{(16\pi^2)^2} \left[(1+2N_c)\cC_{qq}^{(1)} + 3 \cC_{qq}^{(3)} + 2 (1 + N_c) \cC_{tt} - 2N_c \cC_{qt}^{(1)}\right]
\log^2 \left(\frac{\mu^2}{\Lambda^2}\right)  \,.
\label{eq:CHDNLL}
\end{align}
Though this is a 2-loop effect, numerically it behaves more like a 1-loop effect due to the double logarithm and a large anomalous dimension. To gain some confidence in this, let us consider an example where we activate only $\cO_{tt}$ in the UV. We will take the UV scale to be $\Lambda = 25$ TeV and then check the RG-generated value of $\cC_{HD}$ at $m_Z$ via several methods. We find $\cC_{HD}(m_Z)/\cC_{tt}(25~{\rm TeV}) = (7.5, 7.7, 7.9, 8.1)\times 10^{-2}$ using, in order, the analytic expression~\cref{eq:CHDNLL} with $y_t(25~{\rm TeV}) = 0.75$, \texttt{DSixTools} method 1 (full integration), \texttt{DSixTools} method 3 (EMM), and \texttt{wilson} full integration~\cite{Aebischer:2018bkb}. The underestimation of the analytic formula is due to the neglected running of $y_t$ that is taken into account in the other methods. For comparison, the value of $y_t^2 N_c \log(\Lambda^2 / m_Z^2)/16\pi^2 \approx 0.12$, so this 2-loop LL effect has the numerical size of a naively-estimated first LL effect.

To the best of our knowledge, these anomalously large double-log custodial violations were first pointed out by us in~\cite{Allwicher:2023aql}, and further studied in~\cite{Allwicher:2023shc}.\footnote{Another such anomalously large double-log effect was found in a different context, involving the running of some dimension-5 interactions in the EFT of an axion-like particle into $\cO_{HD}$~\cite{Biekotter:2023mpd}.} They were also captured in the top-operator analysis of~\cite{Garosi:2023yxg}, which numerically integrates the full 1-loop RG equations.

\begin{figure}[t]
    \centering
    \begin{tikzpicture}[thick,>=stealth,scale=1]
        \node[wc] at (0,0) {};
        \node[below=7pt] at (0,0) {$\cC_{4q}$};
        \draw[midarrow] (0,0) -- (-1,1) node[left] {$q$};
        \draw[midarrow] (-1,-1) node[left] {$q$} -- (0,0);
        \draw[midarrow] (1,-1) -- (0,0);
        \draw[midarrow] (0,0) -- (1,1);
        \draw[midarrow] (1,1) -- (1,-1);
        \draw[dashed] (1,1) -- (2,1) node[right] {$H$};
        \draw[dashed] (1,-1) -- (2,-1) node[right] {$H$};
        \draw[->] (2.5,0) -- +(0.5,0) node[right] {$\cC_{Hq}$};
        \begin{scope}[xshift=7.5cm]
            \node[wc] at (0,0) {};
            \node[below=7pt] at (0,0) {$\cC_{Hq}$};
            \draw[dashed] (0,0) -- (-1,1) node[left] {$H$};
            \draw[dashed] (-1,-1) node[left] {$H$} -- (0,0);
            \draw[midarrow] (1,-1) -- (0,0);
            \draw[midarrow] (0,0) -- (1,1);
            \draw[midarrow] (1,1) -- (1,-1);
            \draw[dashed] (1,1) -- (2,1) node[right] {$H$};
            \draw[dashed] (1,-1) -- (2,-1) node[right] {$H$};
            \draw[->] (2.5,0) -- +(0.5,0) node[right] {$\cC_{HD}$};
        \end{scope}
    \end{tikzpicture}
    \caption{Two-loop leading-log RG mixing of $\cC_{4q} \rightarrow \cC_{Hq} \rightarrow \cC_{HD}$. See~\S~\ref{sec:NLL} for details.
    \label{fig:NLLrunning}}
\end{figure}
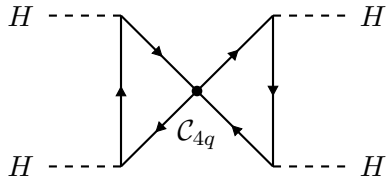
\begin{figure}[t]
    \centering
    \begin{tabular}{c@{\hskip 1cm}c}
    \begin{tikzpicture}[thick,>=stealth]
        \node[wc] at (0,0) {};
        \node[below=7pt] at (0,0) {$\cC_{4q}$};
        \draw[midarrow] (-1,1) -- (0,0);
        \draw[midarrow] (0,0) -- (-1,-1);
        \draw[midarrow] (-1,-1) -- (-1,1);
        \draw[midarrow] (0,0) -- (1,1);
        \draw[midarrow] (1,1) -- (1,-1);
        \draw[midarrow] (1,-1) -- (0,0);
        \draw[dashed] (-1,1) -- (-2,1) node[left] {$H$};
        \draw[dashed] (1,1) -- (2,1) node[right] {$H$};
        \draw[dashed] (-1,-1) -- (-2,-1) node[left] {$H$};
        \draw[dashed] (1,-1) -- (2,-1) node[right] {$H$};
    \end{tikzpicture}
    \end{tabular}
    \caption{Two-loop contribution of $\cC_{4q}$ to $\cC_{HD}$. It has the structure given in~\cref{eq:CHD2loop}, including the double log from the 1-loop RGE as well as a single log from 2-loop RGE.}
    \label{fig:twoLoopCHD}
\end{figure}

Since this is a 2-loop effect obtained from resumming the 1-loop RGE, the reader might wonder if the 2-loop RGE can change our conclusions. While the 2-loop anomalous dimension matrix in the SMEFT is not yet known, we can estimate its impact via the following argument. Let us define $\alpha_t = y_t^2 / (4\pi)$. Then one-loop RGE effects take the form $[\alpha_t \log(\mu^2/\Lambda^2)]^n$, while two-loop contributions would behave as $\alpha_t [\alpha_t \log(\mu^2/\Lambda^2)]^n$. Therefore, in general we expect the full $O(\alpha_t^2)$ contribution to $\cC_{HD}$ to be a second-order logarithmic polynomial, \eg for $\cC_{tt}$ we would write
\begin{equation}
[\cC_{HD}]_{\text{2-loop}} = \frac{N_c (N_c+1)}{4\pi^2} \alpha_t^2 \bigg[ \underbrace{ \log^2(\mu^2/\Lambda^2)}_{\text{1-loop RGE}} + \underbrace{c_1 \log(\mu^2/\Lambda^2)}_{\text{2-loop RGE}} + \underbrace{c_2}_{\text{finite}} \bigg] \cC_{tt}\,,
\label{eq:CHD2loop}
\end{equation}
where $c_{1,2}$ are constants unknown to the 1-loop RGE, presumably of $O(1)$. This formula corresponds to computing 2-loop graphs such as~\cref{fig:twoLoopCHD}, some of which should factorize into the product of the two 1-loop graphs in~\cref{fig:NLLrunning}.
The key takeaway is that the 2-loop RG correction is expected to come proportional to a single logarithm. For $\mu = m_Z$ and $\Lambda \simeq 2.5$ TeV, the two-loop RGE and finite parts could change our results by around 15\% and 2\%, respectively. However, it is clear that the 1-loop RGE dominates over the 2-loop and any finite parts in the large mass gap limit. For example, for $\Lambda \simeq$ 25 TeV we estimate that the 2-loop RGE could give a 9\% correction, while finite contributions would be $<1$\%.

During the preparation of this article, we became aware of a very interesting ongoing effort to perform an $O(\alpha_t^2)$ 2-loop computation of some EW precision observables~\cite{uliANDluc}. Their computation of the $T$ parameter indeed has the form as~\cref{eq:CHD2loop}. In particular, it includes a double-log contribution from third-family 4-quark operators, which we have checked has the same coefficient as~\cref{eq:CHDNLL}, as necessary for the $\mu$-independence of the computation. Furthermore, they find that the single log has a coefficient $c_1 = 1/2$, which gives the 2-loop anomalous dimension for direct $\cC_{tt} \rightarrow \cC_{HD}$ mixing, while the finite part $c_2 = 0$~\cite{uliANDlucPC}. Thus, the impact of the single log from the 2-loop RGE is about half our estimate, and we incorporate it in all numerical results using~\cref{eq:CHD2loop} (we resum the double-log using the 1-loop RG equations). In the calculation of Ref.~\cite{uliANDlucPC}, the 2-loop contribution of $\cC_{tt}$ to $T$ vanishes for $\mu = m_t$. The same is true for some $O(\alpha_t)$ contributions in~\cite{Allwicher:2023aql}. Given the scale sensitivity of the double-log term, this suggests that the error we make by only performing a tree-level computation of the EW observables is minimized if we adopt this choice, so we take $\mu_{\rm EW} = m_t$ for all our numerical results.

\subsubsection{Comparison with high-energy bounds from top production}
Importantly, Refs.~\cite{Dawson:2022bxd,Allwicher:2023shc,Garosi:2023yxg,Bellafronte:2023amz,uliANDlucPC} agree that the loop effects described in~\S~\ref{sec:LLrunningqq} and~\S~\ref{sec:NLL} give stronger bounds on third-family 4-quark operators than current high-energy LHC constraints from top production. That is to say, current EW precision data gives the best bound on these operators, not high-energy probes as one would naively expect via the argument that 4-fermion interactions grow as $E^2$. There are three basic reasons for this: i) 4-top operators can only be probed at tree-level in $pp \rightarrow \bar t t \bar t t$ processes, which currently have considerable uncertainties, ii) 4-top operators mainly contribute to 2-top production at one loop, since bottom-quark initiated processes are strongly PDF suppressed, and iii) though the contribution of 4-top operators to EW observables is also loop suppressed, the precision at which they are measured compensates for this. 

Using $\cO_{tt}$ as an example, the latest 4-top experimental analysis from ATLAS finds a lower bound of about 725 GeV on $\cC_{tt}$~\cite{ATLAS:2023ajo}, while 
a combined analysis of top-quark production at the LHC finds a bound of about 900 GeV~\cite{Ethier:2021bye}. Finally, from the 2-loop contribution to $T$ we get a bound of 1.1 TeV, as shown in~\cref{tab:currentBounds}.  High-energy bounds on 4-top operators are expected to improve during the HL-LHC phase, presumably by about a factor $\sqrt{2}$ on the scale~\cite{Degrande:2024mbg,Celada:2024mcf}, which could make them competitive. As we will show next, however, the current bound on $\cO_{tt}$ from EW data is already strong enough to give the dominant constraint on the natural parameter space of the right compositeness scenario.

%----------------------------------------------------------------------------
\section{Probing composite Higgs models}
\label{sect:CHmodels}

We now apply the results of the previous sections to constrain the SILH+TQ scenario. The scaling rules of~\cref{eq:scalingEFT} allow us to do the matching for all WCs up to model-dependent $O(1)$ factors. One option is to attempt some kind of global fit, which would lead to the most conservative bounds by exploiting any complicated flat directions. Since we are skeptical that realistic models would populate such flat directions, we opt here for a mixed approach in the style of Ref.~\cite{Glioti:2024hye}. For each WC in~\cref{tab:boundTable}, we perform the matching according to~\cref{eq:Umatching,eq:NUmatching}, set the $O(1)$ factors to 1, and require that the 95\%~CL single operator bound be satisfied. This will allow us to overlay all of the constraints together in the $(m_*, g_*)$ plane, fully capturing the parametric scaling of each operator with $g_*$, $m_*$, and $\eps_{L,R}$. Due to the $O(1)$ uncertainty, these regions should not be taken as absolute, but rather as representative bounds on a generic SILH-TQ scenario. It may happen that a particular model beats one or another bound, but then the next strongest constraint can easily be read from the plots. As we will see, a compilation of overlapping constraints will lead to what we think is a clear picture on the overall lower bound on $m_*$.

Our default will be to study ``symmetry-protected models" where we assume that the strong sector respects custodial and $P_{LR}$ symmetry. In practice, what this means for us is that we take $\cC_T = \cC_{Ht} = 0$ and $\cC_{Hq}^{(1)} = -\cC_{Hq}^{(3)}$ at tree level, which avoids the top four bounds in~\cref{tab:currentBounds}.
We will also compare our results to a scenario with no $P_{LR}$ protection, where we assume $\cC_{Hq}^{(1)}$, $\cC_{Hq}^{(3)}$, and $\cC_{Ht}$ are all present at tree level. In this case, there is also an important flavor constraint from $B_s \rightarrow \mu^+\mu^-$ that we include (see~\S~\ref{sec:FVops}), which has the same parametric scaling as $Z b_L \bar  b_L$ vertex corrections. The stronger bound currently comes from $B_s \rightarrow \mu^+\mu^-$, but this will change at FCC-ee, so we keep both.

%----------------------------------------------------------------------------
\subsection{Main results}
\label{sec:mainResults}
Our parameters are $g_*$, $m_*$, and due to the top Yukawa constraint, only one of the mixing parameters $\eps_{L,R}$. Accordingly, we display our main results as plots in $(m_*, g_*)$ plane for three different mixing choices: left compositeness (\cref{fig:LeftCompPlots}), mixed compositeness (\cref{fig:MixedCompPlots}), and right compositeness (\cref{fig:RightCompPlots}). The definition of these scenarios in terms of the mixing parameters was given in~\S~\ref{sec:SITQops}.

In each mixing scenario, we show constraints from three classes of operators: i) Flavor-universal operators, shown on the plots as a gray region labeled ``Universal", ii) Flavor-violating operators, shown as a sand-colored region labeled ``Flavor", and iii) the leading constraints from RGE of flavor non-universal operators. For comparison, we also show constraints (purple region) from finite contributions to the $T$ parameter that we estimated in~\cref{eq:CHDfinite}. Finally, the region to the left of the dashed-black line would be excluded if no $P_{LR}$ symmetry was invoked. Otherwise, the full white region is allowed. Some general comments are now in order.
\begin{itemize}
\item For the flavor-universal constraints shown in all current data plots, we use the exclusion region that was derived in Ref.~\cite{Glioti:2024hye}. The reason is that our EW fit reproduces their bound on the $S$ parameter, while the two other important operators $\cO_H$ and $\cO_{2B,2W}$ are currently better constrained by LHC data. On the other hand, the flavor-universal constraints shown in all FCC-ee plots come entirely from our projected EW likelihood, assuming the observable improvements in~\cref{sec:appFCC}. 
\item The sand-colored ``Flavor" exclusion region comes from the bound on $B$-meson mixing discussed in~\S~\ref{sec:FVops}. In the future projection plots, we assume that the bound on the Wilson coefficient will improve by a factor of 4~\cite{Cerri:2018ypt,Charles:2020dfl}. Since $V_{cb}$ and hadronic uncertainties are currently the limiting factor, the expected improvement at HL-LHC and FCC-ee is approximately the same.
\item Since there are many flavor non-universal operators, we plot only those with the strongest bounds (after assuming a loop suppression for dipoles as discussed in~\S~\ref{sec:SITQops}). Explicitly, the SITQ operators we plot are: $\cC_{Hq}^{(-)}$, $\cC_{tt}$, $\cC_{qq}^{(1,3)}$, $\cC_{qt}^{(1)}$, $\cC_{qD}^{(1,3)}$, and $\cC_{tD}$. We have grouped $\cC_{qq}^{(1,3)}$ and $\cC_{qD}^{(1,3)}$ together since they have the same parametric scaling- we simply display the strongest bound in each case.
\end{itemize}
We begin discussing our main results with the flavor-universal constraints, which do not depend on the mixing choice. As can be seen from~\cref{eq:Umatching}, $\cC_H \propto g_*^2/m_*^2$ dominates for large $g_*$, $\cC_{W,B} \propto 1/m_*^2$ (corresponding to the $S$ parameter) gives a $g_*$-independent bound, while $\cC_{2W,2B} \propto 1/(g_* m_*)^2$ ($W$ and $Y$ parameters) give the dominant bound for small $g_*$. This explains the shape of the gray universal exclusion region in~\cref{fig:LeftCompPlots,fig:MixedCompPlots,fig:RightCompPlots}. 

As previously mentioned, $\cC_H$ and $\cC_{2B,2W}$ are currently better constrained by LHC data, while the best bound on $\cC_{W,B}$ comes from EW precision data. This situation is projected to change at FCC-ee, where the best bound on all the universal operators will come from EW precision data. While this is clear for $\cC_{W,B}$ and $\cC_{2W,2B}$ since they can be seen to directly affect $W$- and $Z$-pole observables after using the EOM, the projected bound on $\cC_H$ of 6 TeV shown in~\cref{tab:FccBounds} is coming from RGE. According to the map in~\cref{app:A}, the SILH operator $\cO_H$ corresponds to $\cO_{H\Box}$ in the Warsaw basis. From Ref.~\cite{Alonso:2013hga}, we see there is first LL running in $g_1$ of $\cO_{H\Box} \rightarrow \cO_{HD}$, again with a surprisingly large anomalous dimension $\propto 80/3$. There is also $g_2$ running of $\cO_{H\Box} \rightarrow [\cO_{Hl}^{(3)}]_{U}$, where $U$ indicates a universal contribution. Both of these RG-generated operators affect $m_W$, which indeed is the constraining observable for $\cO_H$ according to~\cref{tab:FccBounds}. The former effect is due to $g_1$-induced custodial violations, while the latter affects the extraction of $G_F$. We note that this RG-induced bound on $\cO_H$ is significantly stronger than the projected $\simeq 1.1$ TeV at the HL-LHC~\cite{Glioti:2024hye}. 

Moving on to the non-universal and flavor-violating constraints, we will now make some comments on each mixing scenario individually.
\subsubsection{Left compositeness}
\begin{figure}[t]
    \centering
    \subfloat[\centering Left compositeness (current)]{{\includegraphics[width=0.485\textwidth]{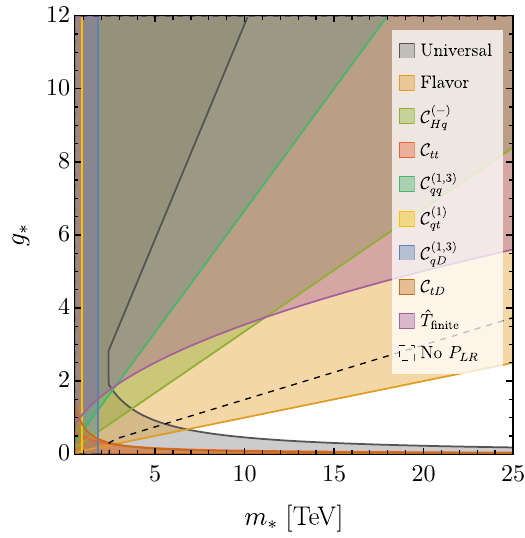} }}%
    \hspace{2.35mm}
    \subfloat[\centering Left compositeness (FCC-ee)]{{\includegraphics[width=0.475\textwidth]{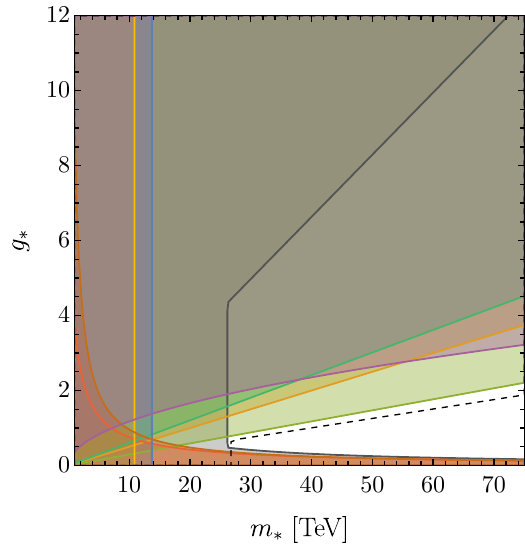} }}%
    \caption{Constraints on the left compositeness scenario. The legend is the same for both panels, and the white region is allowed. For current (future) data we take $\Lambda = 2.5$ TeV (25 TeV).}%
    \label{fig:LeftCompPlots}%
\end{figure}
Our results for the left compositeness case corresponding to a fully composite $q_L^3$ are shown in~\cref{fig:LeftCompPlots}. Starting with current data on the left side, we see that the strongest bound for large $g_*$ comes from flavor violation due to $B$-meson mixing, while for small $g_*$ the bound is given by universal constraints from the operators $\cO_{2B,2W}$. The overlap of the two constraints currently gives a naive lower bound of $m_* \gtrsim 7$ TeV. 

On the other hand, our projection for FCC-ee in the right panel shows the dominant bound for large $g_*$ will instead be given by the non-universal operator $\cC_{Hq}^{(-)}$ (green region) running into the $T$ parameter. As discussed in~\S~\ref{sec:LLrunningHq}, the log-enhanced RGE dominates over estimations for finite contributions to $T$ shown in purple. The bound for small $g_*$ is still given by $\cO_{2B,2W}$, while the $S$ parameter sets an overall lower bound of $m_* \gtrsim 26$ TeV.

We also see that the situation doesn't change much if one does not invoke $P_{LR}$, which would exclude the region to the left of the dashed-black line. Currently, this would just introduce another flavor bound from $B_s \rightarrow \mu^+ \mu^-$ that is similar to the $B$-meson mixing bound (both scaling as $g_*^2/m_*^2$ in this case). At FCC-ee, the dominant bound would come from tree-level $Z b_L \bar b_L$ vertex corrections, with the first LL running $\cO_{Ht} \rightarrow \cO_{HD}$ setting an overall lower bound of $m_* \gtrsim 26$ TeV (similar to $S$).

Finally, as can be seen in~\cref{fig:LeftCompPlots} and~\cref{app:C}, we note that the left compositeness scenario is quite unnatural due to the expectation of large flavor and loop-level custodial violations. This can be seen clearly in~\cref{fig:LeftCompNatPlots}, which shows that the current lower bound on $f = m_*/g_*$ is about 10 TeV, while FCC-ee is projected to yield $f\gtrsim 35$ TeV.

\subsubsection{Mixed compositeness}
\begin{figure}[t]%
    \centering
    \subfloat[\centering Mixed compositeness (current)]{{\includegraphics[width=0.485\textwidth]{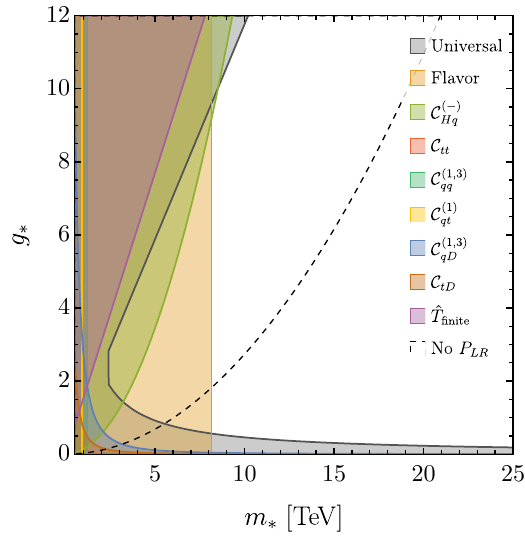} }}%
    \hspace{2.35mm}
    \subfloat[\centering Mixed compositeness (FCC-ee)]{{\includegraphics[width=0.475\textwidth]{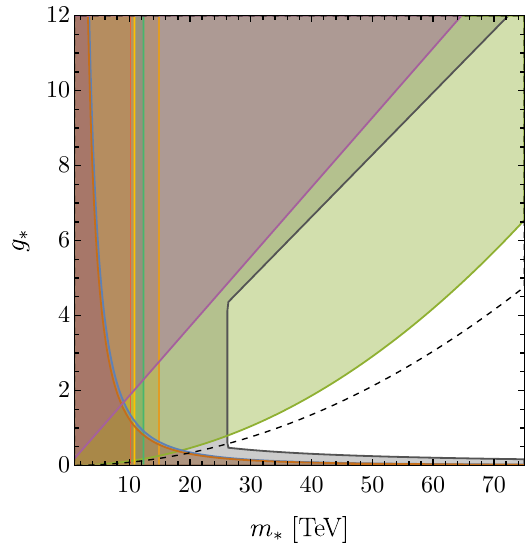} }}%
    \caption{Constraints on the mixed compositeness scenario. The legend is the same for both panels, and the white region is allowed. For current (future) data we take $\Lambda = 2.5$ TeV (25 TeV).}%
    \label{fig:MixedCompPlots}%
\end{figure}
We move now to the mixed compositeness case, corresponding to equally composite $q_L^3$ and $t_R$. The results are shown in~\cref{fig:MixedCompPlots}. Starting with current data on the left side, we see that $B$-meson mixing now gives a $g_*$-independent bound of $m_* \gtrsim 8$ TeV. For large $g_*$, the running $\cO_{Hq}^{(-)} \rightarrow \cO_{HD}$ (green region) and the universal bound from $\mathcal{O}_H$ are the most important. For small $g_*$, the bound is again given by universal constraints from the operators $\cO_{2B,2W}$.

Our projection for FCC-ee in the right panel of~\cref{fig:MixedCompPlots} shows the dominant bound for large $g_*$ is again given by the running $\cO_{Hq}^{(-)} \rightarrow \cO_{HD}$ (green region). Here, finite contributions $T$ are more suppressed due to smaller $\lambda_L$ than in the left compositeness scenario. The bound for small $g_*$ is given by $\cO_{2B,2W}$, while the $S$ parameter again sets an overall lower bound of $m_* \gtrsim 26$ TeV

We see that there are more consequences to not invoking $P_{LR}$ in the mixed compositeness case, which would exclude the region to the left of the dashed-black line. The main reason for current data is that $\Delta F=1$ processes are now enhanced by one power of $g_*$ compared with $\Delta F=2$. Consequently, the bound from $B_s \rightarrow \mu^+ \mu^-$ dominates over the one from $B$-meson mixing for $g_* \gtrsim 1$. At FCC-ee, the dominant bound in the no $P_{LR}$ case would again come from tree-level $Z b_L \bar b_L$ vertex corrections, pushing the viable parameter space to smaller $g_*$.

As can be seen in~\cref{fig:MixedCompPlots} and~\cref{app:C}, the mixed compositeness scenario is more natural than the left compositeness case, even though the lower bounds on $m_*$ are relatively similar. This is because it is possible to go to very large values of $g_*$ in the mixed case. This can be seen clearly in~\cref{fig:MixedCompNatPlots}, which shows that the current lower bound on $f = m_*/g_*$ is only about 1 TeV, while FCC-ee is projected to yield $f\gtrsim 15$ TeV. We note, however, that to have $f\sim 1$ TeV, a coupling $g_* \gtrsim 8$ is required.

\subsubsection{Right compositeness}
\label{sec:Rcomp}
\begin{figure}[t]
    \centering
    \subfloat[\centering Right compositeness (current)]{{\includegraphics[width=0.485\textwidth]{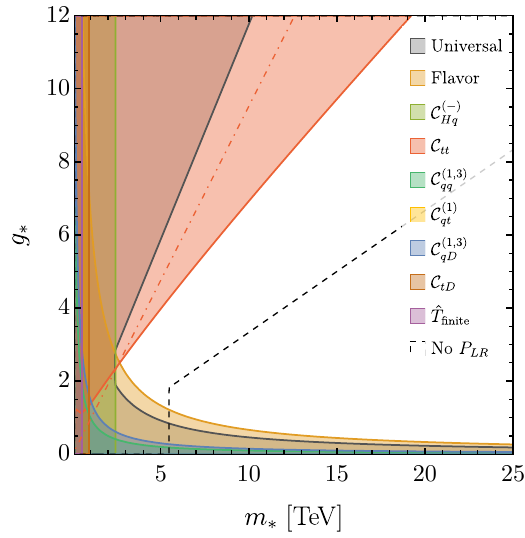} }}%
    \hspace{2.35mm}
    \subfloat[\centering Right compositeness (FCC-ee)]{{\includegraphics[width=0.475\textwidth]{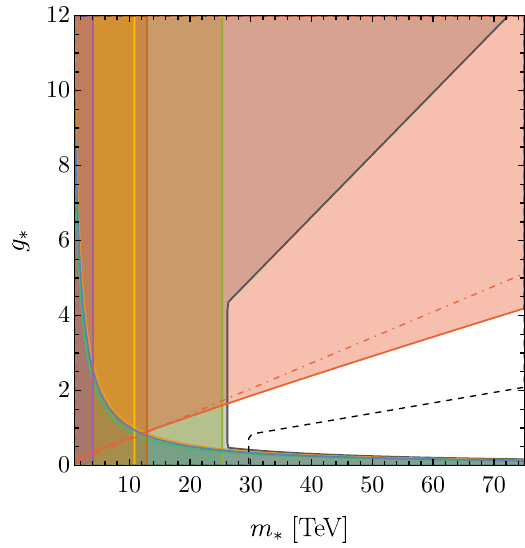} }}%
    \caption{Constraints on the right compositeness scenario. The legend is the same for both panels, and the white region is allowed. The difference between the solid and dot-dashed red lines is explained in~\S~\ref{sec:Rcomp}. For current (future) data we take $\Lambda = 2.5$ TeV (25 TeV).}%
    \label{fig:RightCompPlots}%
\end{figure}
Finally, we discuss the right compositeness case corresponding to a fully composite $t_R$. This is perhaps the most interesting case, with the results shown in~\cref{fig:RightCompPlots}. Starting with current data on the left side, we see that $B$-meson mixing, now scaling as $1/(g_* m_*)^2$, is giving the best bound for small $g_*$. For large $g_*$, we have a novel bound coming from RG custodial violations due to the 2-loop double-log mixing $\cO_{tt} \rightarrow \cO_{Ht} \rightarrow \cO_{HD}$ (red region) that was discussed in detail in~\S~\ref{sec:NLL}. We show this bound in two ways i) fixing $\Lambda$, solving the full 1-loop RG equations numerically (same as the other bounds) while taking into account the single log from the 2-loop RGE in~\cref{eq:CHD2loop} analytically (dot-dashed red), or ii) Using~\cref{eq:CHD2loop} directly without resummation, taking $\Lambda = m_*$ and using $y_t(m_*)$ (solid red). The first approach has the advantage of resumming the double log and correctly taking into account the running of $y_t$. The second approach has the advantage of allowing us to continuously renormalize for every value of $m_*$, which fully captures the high sensitivity of the double log to heavy NP. We believe the second approach is more accurate, which is why we shade that region.

The point where the bound due to $\cO_{tt}$ crosses with the flavor one corresponds a lower bound $m_* \gtrsim 3$ TeV, which is a bit stronger than what was found in the tree-level only analysis of~\cite{Glioti:2024hye}. It is interesting that flavor-universal effects are currently subleading to flavor non-universal and flavor-violating probes in the right compositeness case.

The FCC-ee projection in the right panel of~\cref{fig:RightCompPlots} shows that the double-log RG mixing $\cO_{tt} \rightarrow \cO_{Ht} \rightarrow \cO_{HD}$ (red region) continues to be the dominant probe in the large $g_*$ region. Here, finite contributions to $T$ are the most suppressed as it is the case with the smallest $\lambda_L$. The bound for small $g_*$ is given by several similarly-sized effects: the usual flavor-universal bound due to $\cO_{2B,2W}$, the bound from $B$-meson mixing, and RGE bounds from the non-universal operators $\cO_{qD}^{(1,3)}$ and $\cO_{qq}^{(1,3)}$, the latter of which is another double-log effect. Finally, the $S$ parameter and the first LL running $\cO_{Hq}^{(-)} \rightarrow \cO_{HD}$ (green region) tell a similar story of an overall lower bound $m_* \gtrsim 26$ TeV.

The right compositeness case has the most severe consequences if $P_{LR}$ is not invoked, which would exclude the region to the left of the dashed-black line. This is due to the aforementioned $B_s \rightarrow \mu^+ \mu^-$ (current) and $Z b_L \bar b_L$ (FCC-ee) constraints, which are now $g_*$-independent, pushing up the overall scale. The biggest change without $P_{LR}$, however, comes from $\cC_{Ht} \propto \eps_R^2/f^2$ at tree-level, which is unsuppressed in the right compositeness case. This operator then yields 1-loop custodial violations due to the first LL running $\cO_{Ht} \rightarrow \cO_{HD}$, pushing the viable parameter space to smaller values of $g_*/m_*$.

As can be seen in~\cref{fig:RightCompPlots} and~\cref{app:C}, the right compositeness scenario is similarly natural to the mixed compositeness case, but it allows for a much wider range of couplings $g_*$. This is because the right compositeness case allows for the lowest overall mass scale $m_*$. This can be seen clearly in~\cref{fig:RightCompNatPlots}, where the current lower bound on $f = m_*/g_*$ is about 1-1.5 TeV, and FCC-ee is projected to yield $f\gtrsim 16$-18 TeV. Unlike the mixed compositeness case, these 1 TeV values for $f$ can be obtained for any value of the coupling $g_* \gtrsim 2$. Right compositeness therefore allows the most flexibility in $g_*$, which is important for obtaining acceptable values for the Higgs potential parameters, where the optimal value seems to be around $g_* = 2$-3~\cite{Giudice:2007fh,Bellazzini:2014yua,Fuentes-Martin:2022xnb}.

\subsubsection{RG custodial violations and $m_W$}
\label{sec:RGCVandMW}
We have seen that custodial violations due to both 1- and 2-loop effects are quite generic in composite Higgs models and provide powerful constraints on the parameter space. Since these effects are due to SM custodial violations $y_t \gg y_b$, they are unavoidably present even in models where the strong sector respects custodial symmetry. Because $m_Z$ is fixed in our EW input scheme, these violations manifest as shifts in the $W$-boson mass~\cite{Berthier:2015oma}
\begin{equation}
\frac{\delta m_W}{m_W} = -\frac{v^2}{4}\frac{\cC_{HD}}{(1-t_W^2)} = \frac{\hat T}{2(1-t_W^2)}\,,
\end{equation}
where $t_W = g_1/g_2$. Thus, positive shifts in $\hat T$ correspond to an increase in the $W$ mass. This is interesting as the measured value of $m_W$ is roughly $\sim 2\sigma$ larger than the SM theory prediction, see \eg\cite{deBlas:2021wap}.
 The largest contributions are first LL effects due to $\cC_{Hq}^{(1)}$ and $\cC_{Ht}$ shown in~\cref{eq:CHDLL}, and 2-loop double-log effects from the 4-top operators in~\cref{eq:CHDNLL}. The latter effects always give positive contributions to $\hat T$ if the WCs are negative,\footnote{The exception is $\cC_{qt}^{(1)}$, but it is suppressed by $g_*^2$ compared to the others, see~\cref{eq:NUmatching}.} which is indeed what one obtains by integrating out a vector singlet or color octet at tree level. The first LL effects give positive contributions to $\hat T$ if $\cC_{Hq}^{(1)}$ is positive, or if $\cC_{Ht}$ is negative. According to~\cite{deBlas:2017xtg}, the top partners that couple linearly to the SM and can generate these operators are $Q_1$, $Q_7$, $U$, $T_1$, and $T_2$. A positive contribution to $\hat T$ is generated for $Q_1 \sim ({\bf 3},{\bf 2})_{1/6}$, $U \sim ({\bf 3},{\bf 1})_{2/3}$, and $T_2 \sim ({\bf 3},{\bf 3})_{2/3}$. If $P_{LR}$ symmetry is invoked, $Q_1$ would be accompanied by $Q_7\sim ({\bf 3},{\bf 2})_{7/6}$ such their tree-level contribution to $\cC_{Ht} $ vanishes, and $T_2$ would come with $T_1\sim ({\bf 3},{\bf 3})_{-1/3}$, yielding $\cC_{Hq}^{(1)}=\cC_{Hq}^{(3)} = 0$. Finally, $U$, a heavy partner of $t_R$, respects $P_{LR}$ since it is a custodial singlet, yielding $\cC_{Hq}^{(1)}=-\cC_{Hq}^{(3)}$ and giving a positive $\delta m_W$. 

 We thus conclude that it is rather generic to have positive contributions to the $W$ mass via RG-induced custodial violations, as was also pointed out in~\cite{Crivellin:2022fdf,Allwicher:2023aql}. This is particularly true in the right compositeness case, where RG custodial violations are giving the dominant bounds, so one could realistically have sizable positive shifts in $m_W$ without being in conflict with other constraints.

\subsection{Future summary plot}
\label{sec:futurePlot}
\begin{figure}[t]%
    \centering
    \subfloat[\centering Left compositeness]{{\includegraphics[width=0.31\textwidth]{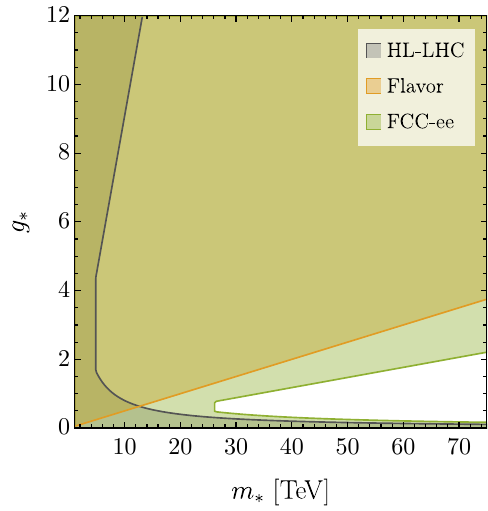} }}%
    \hspace{1mm}
    \subfloat[\centering Mixed compositeness]{{\includegraphics[width=0.31\textwidth]{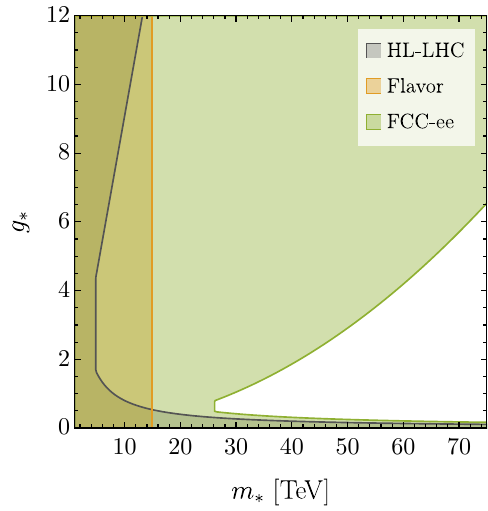} }}%
    \hspace{1mm}
    \subfloat[\centering Right compositeness]{{\includegraphics[width=0.31\textwidth]{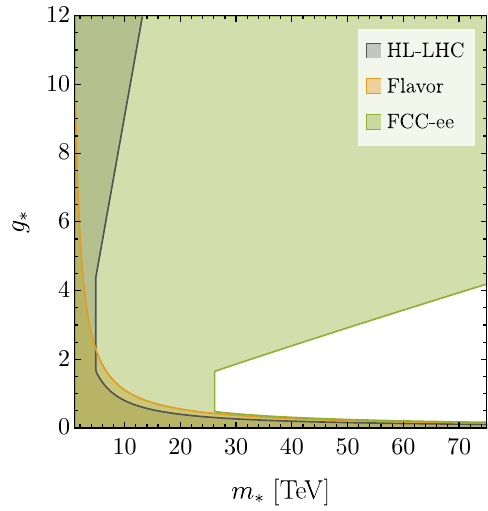} }}%
    \caption{Summary plot of future projections by sector. We assume $P_{LR}$ symmetry for the FCC-ee projection, and $\Lambda = $ 25 TeV. The white region is allowed.}%
    \label{fig:summaryPlot}%
\end{figure}

Having discussed individual bounds in detail in~\S~\ref{sec:mainResults}, we now want to provide some simplified summary plots to clearly demonstrate the potential reach of a future tera-$Z$ machine like FCC-ee. To this end, we show in~\cref{fig:summaryPlot} the overall projected FCC-ee exclusion region for each mixing scenario (green), assuming $P_{LR}$ symmetry. That is, we take the bounds outlining the white region in~\cref{fig:LeftCompPlots,fig:MixedCompPlots,fig:RightCompPlots}, assuming $P_{LR}$ in each case. Also shown are the regions expected to be probed with improvements in flavor (sand-colored) and LHC bounds (gray) over the next 20 years, where we have used the HL-LHC projections of~\cite{Glioti:2024hye}.

Let us re-emphasize that this is not a global fit, which we leave for future work. While we expect the scaling of the bounds is correct, they of course come with an $O(1)$ uncertainty. However, the overall takeaway message is that FCC-ee is capable of setting a lower bound of $m_*\gtrsim 25$ TeV, even if the model is protected by $P_{LR}$ symmetry. This statement is independent of the mixing choice for the SITQ. The large $g_*$ region is always dominantly constrained by non-universal RG effects, while the small $g_*$ region is better probed by universal effects (at both the HL-LHC and FCC-ee). Flavor bounds due to $B$-meson mixing play a complementary role, with the exclusion region depending strongly on the mixing choice. We note that FCC-ee will be able to add the most new information in the mixed and right compositeness cases, which are the most natural. 

Compared to Figure 7 of~\cite{deBlas:2019rxi}, which does not consider a SITQ, we find generically stronger bounds. This is especially true in the large $g_*$ region, where flavor non-universal RG effects play a dominant role. To the best of our knowledge, Ref.~\cite{deBlas:2019rxi} matches their WCs to composite Higgs model parameters at the EW scale, which neglects RGE and operator mixing. They also assume some intrinsic theory uncertainty, which is the reason \eg for their weaker bound on the $S$ parameter. However, as can be seen clearly in~\cref{fig:summaryPlotTL}, the stronger bounds we find here are dominantly coming from RG mixing effects that were not taken into account in Ref.~\cite{deBlas:2019rxi}. 

We also find stronger bounds than the best limits from $ee\rightarrow tt$ at a 3 TeV center-of-mass CLIC. In particular, Ref.~\cite{Banelli:2020iau} finds a bound of $f \gtrsim 7.7$ TeV from considering first LL running in $g_1$ of the operator $\cO_{tt} \rightarrow \cO_{tD}$, while we obtain $f \gtrsim 15.7$ TeV from the double-log mixing in $y_t$ of $\cO_{tt} \rightarrow \cO_{HD}$ discussed in~\S~\ref{sec:NLL}. Ref.~\cite{Durieux:2018ekg} instead obtains a relatively $g_*$-independent bound of $m_* \gtrsim 30$ TeV at 68\% CL by considering tree-level matching contributions to $\cO_{\psi D}$ operators. Extrapolating to the 95\% CL exclusion gives a bound comparable to our bounds on $\cC_{\psi D}$, which are complementary but subleading constraints.
%

%-----------------------------------------------------------------------------
\section{Discussion and conclusions} 
\label{sec:conclusions}
%-----------------------------------------------------------------------------
In this work, we performed a comprehensive analysis of the leading loop-level phenomenology of composite Higgs models. We did this in the context of an EFT where the light degrees of freedom are the SM gauge sector, a strongly-interacting light Higgs, and a strongly-interacting top quark. The model-independent loop contributions are taken into account by using the full RG equations to evolve operators expected to be generated at tree level in composite Higgs models from the compositeness scale $\Lambda \sim m_*$ down to the EW scale, where precision observables are computed. This procedure resums any large logarithms and fully captures operator mixing beyond the first leading-logarithmic approximation. In doing so, we find that several flavor non-universal operators generated due to the SITQ have large RG mixings into operators contributing to EW precision observables. In the universal case, we also point out a sizable RG mixing of $\cO_{H\Box} \rightarrow \cO_{HD}$ that will allow FCC-ee to provide a strong bound on the SILH operator $\cO_H$ via a measurement of $m_W$. In~\S~\ref{sec:RGEeffects}, we argued that RG contributions are more important than estimates for model-dependent finite matching contributions that are frequently found in the composite Higgs literature.

We account for the impact of flavor non-universal effects by examining three options for the top mixing: fully composite $q_L^3$, equally composite $q_L^3$ and $t_R$, and fully composite $t_R$. As expected, the left compositeness scenario is the least natural due to the expectation of large flavor and custodial violations. On the other hand, both the mixed and right compositeness cases currently allow for $f \sim 1$ TeV. The required value of $g_*$, however, is quite different in the two cases: mixed compositeness requires a strongly-coupled theory with $g_* \gtrsim 8$, while right compositeness only requires $g_* \gtrsim 2$. The latter therefore allows for more flexibility in obtaining the correct Higgs potential parameters.

We find that the fully composite $t_R$ scenario allows for the lowest overall scale $m_* \gtrsim 3$ TeV. Interestingly, this case is dominantly constrained by flavor and RG custodial violations due to the non-universal SITQ operators, rather than by universal effects at tree level. New in this work is that the dominant constraint on the most natural parameter space, namely $g_* \gtrsim 2$, comes from 2-loop custodial violations that can be understood as double-log RG mixing of the 4-top operator $\cO_{tt} = (\bar t_R \gamma_\mu t_R)(\bar t_R \gamma^\mu t_R)$ into $\cO_{HD}$ that corresponds to the $T$-parameter in the SMEFT. To the best of our knowledge, this effect was not previously known.

As a generally applicable statement, we find that 4-top operators are currently better constrained by EW precision data rather than high-energy probes from top production at the LHC. This emphasizes that EW precision data, as well as RGE for comparison at the same scale, should be included in future global analyses of 4-top operators. This analysis will be fully enabled by an upcoming 2-loop calculation of the $T$-parameter~\cite{uliANDlucPC}. It will also be very interesting to see if the HL-LHC can achieve parity with existing precision bounds in the coming years.

Since the dominant theme here is RG-induced custodial violations, let us make a few final comments we think are important: i) While invoking $P_{LR}$ enforces $\cC_{Ht} = 0$ and $\cC_{Hq}^{(1)} = -\cC_{Hq}^{(3)}$ at tree-level, in general this does not require $\cC_{Hq}^{(1)} = 0$. This means that $P_{LR}$ does not protect against all first leading-log RG custodial violations in~\cref{eq:CHDLL}. In general, we find strong constraints due to this effect, shown in green in~\cref{fig:LeftCompPlots,fig:MixedCompPlots,fig:RightCompPlots}. ii) If  $P_{LR}$ is not invoked, the right compositeness scenario is expected to be very strongly constrained by RG custodial violations due $\cC_{Ht} \rightarrow \cC_{HD}$ at first leading log (dashed black line in~\cref{fig:RightCompPlots}). While we are aware that effects i) and ii) have pointed out before (see~\cite{Pomarol:2008bh}), to the best of our knowledge they have not been widely appreciated in the composite Higgs literature.

It cannot be overemphasized that all RG custodial violations discussed here are due to the explicit breaking of custodial symmetry by the SM Yukawa couplings, in particular $y_t \gg y_b$. Thus, these RG effects cannot be switched off by clever model building. In that sense, sizable custodial violations are a generic expectation of composite Higgs models. Interestingly, in~\S~\ref{sec:RGCVandMW} we have shown that these RG custodial violations generally have the right sign to lead to an increase in the $W$-boson mass. This is particularly true in the right compositeness case, where RG custodial violations are giving the dominant bounds. It is therefore realistic to expect that if we were on the verge of discovering this scenario, we may very well only see a positive shift in $m_W$ and nothing else at low energy. We find this point intriguing, given the current mild $\sim 2\sigma$ excess in $m_W$.

We have emphasized the importance of considering RG effects in a SILH+TQ scenario, which already provide significant and previously underappreciated constraints, and will play a dominant role in the future. This is explicitly shown in our future projections, which are summarized in~\cref{fig:summaryPlot}. They show that FCC-ee can set a mixing-independent lower bound $m_*\gtrsim 25$ TeV, even on models protected by $P_{LR}$ symmetry. While the bound on $m_*$ is the same in the three cases, the left scenario is still the most unnatural, requiring $f \gtrsim 35$ TeV, while the mixed and right cases yield $f \gtrsim 15$ TeV (still with large $g_*$ in the mixed case). At FCC-ee, non-universal RG effects due to the SITQ completely dominate, regardless of the mixing choice, due to large logarithms that we resum. According to~\cref{tab:FccBounds}, the most important measurements to realize our bounds are $A_e$ and $m_W$, which will also require an immense theory effort in order to match experimental precision. In summary, a future tera-$Z$ factory such as FCC-ee will probe Higgs compositeness with high precision, either leading to a discovery, or pushing fine-tuning of the EW scale to the $\lesssim 10^{-4}$ level.
%
%-----------------------------------------------------------------------------
\section*{Acknowledgments}
%-----------------------------------------------------------------------------
 A very special thank you goes to Ulrich Haisch and Luc Schnell for very helpful discussions and for sharing a preliminary version of their draft~\cite{uliANDlucPC}.
 We also thank Claudia Cornella, Gauthier Durieux, Javier Fuentes-Mart\'in, Admir Greljo, Gino Isidori, David Marzocca, Matthew McCullough, Giuliano Panico, and especially Javier M. Lizana for useful discussions and/or comments on the manuscript. The work of BAS is supported by the STFC (grant No. ST/X000753/1).

\appendix
\section{SILH to Warsaw basis map}
\label{app:A}
The conversion from the SILH to the Warsaw basis is given in~\cite{Alonso:2013hga,Wells:2015uba}. We report it here for our operators as normalized in~\cref{tab:UniversalOps,tab:nonUniversalOps,tab:UniversalOpsLoop}. The gauge equations of motion are
\begin{align}
\partial^{\nu} B_{\mu\nu} &= g_1 j_{1\mu} + g_1 \hyp_h (i H^\dagger  \overleftrightarrow {D_\mu} H) \,, &  j_{1\mu} &= \sum_{\psi = q,l,u,d,e} \hyp_\psi (\bar \psi \gamma_\mu \psi) \,,\\
D^{\nu} W_{\mu\nu}^I &= g_{2} j_{2\mu}^I + \frac{g_2}{2} (i H^\dagger \overleftrightarrow {D_\mu^I} H)\,, & j_{2\mu}^I &= \frac{1}{2}\sum_{\psi = q,l}  (\bar \psi \tau^I \gamma_\mu \psi) \,, \\
D^{\nu} G_{\mu\nu}^A &= g_{3} j_{3\mu}^A \,, & \hspace{18mm} j_{3\mu}^A &= \sum_{\psi = q,u,d}  (\bar \psi T^A \gamma_\mu \psi) \,,
\end{align}
where $\hyp_i$ is the corresponding hypercharge in the normalization where $\hyp_h = 1/2$. Our generators are $T^A = \lambda^A / 2$ and $\tau^I$, where $\lambda^A$ and $\tau^I$ are the Gell-Mann and Pauli matrices, respectively. 
\begin{table}[h]
\centering
\begingroup
\renewcommand*{\arraystretch}{1.5}
\begin{tabular}{|c|c|}
 \hline
 \multicolumn{2}{|c|}{\textbf{Additional universal operators}} \\ \hline \hline
  $\mathcal{O}_{HW} = ig_2 (D^\mu H)^\dagger \tau^I (D^\nu H) W^I_{\mu\nu}$ & $\mathcal{O}_{HB} = ig_1 (D^\mu H)^\dagger (D^\nu H) B_{\mu\nu}$ \\ \hline 
 $\mathcal{O}_g = g_3^2 (H^{\dagger} H) G_{\mu\nu}^AG^{A\,\mu\nu}$ & $\mathcal{O}_\gamma = g_1^2 (H^{\dagger} H) B_{\mu\nu}B^{\mu\nu}$ \\ \hline
  $\mathcal{O}_{3W} = g_2^3 \eps_{IJK} W_{\mu}^{I\nu} W_{\nu}^{J\,\rho} W_{\rho}^{K\,\mu}$ & $\mathcal{O}_{3G} = g_3^3 f_{ABC} G_{\mu}^{A\nu} G_{\nu}^{B\,\rho} G_{\rho}^{C\,\mu}$ \\ \hline
 \multicolumn{2}{|c|}{  $\mathcal{O}_{2G} = -\frac{g_3^2}{2} (D^{\mu}G_{\mu \nu}^A) (D_{\rho}G^{A \rho \nu})$ } \\ 
 \hline
\end{tabular}
\endgroup
\caption{List of additional flavor-universal dimension-six operators. All except $\mathcal{O}_{2G}$ are expected to be generated at one loop in minimally-coupled models.}
\label{tab:UniversalOpsLoop}
\end{table}
Using the equations of motion as well as integration by parts, we find the map for the universal contributions to be
\begin{align}
\cO_H &= -\frac{1}{2} Q_{H \Box} \,, \nn
\cO_T &= - \frac{1}{2}Q_{H \Box} - 2 Q_{H D} \,, \nn 
\cO_{B}  &=
\frac{1}{2} \hyp_h  g_1^2 Q_{H \Box} +2g_1^2 \hyp_h Q_{H D}  + \frac{1}{2}  g_1^2\left[\hyp_l Q_{\substack{H l \\ tt}}^{(1)} +\hyp_e Q_{\substack{H e \\ tt}} + \hyp_q Q_{\substack{H q \\ tt}}^{(1)}+\hyp_u Q_{\substack{H u \\ tt}}+ \hyp_d Q_{\substack{H d \\ tt}} \right],\nn
\cO_{W} &= \frac34 g_2^2 Q_{H \Box} + \frac{g_2^2}{4}\left[Q_{\substack{H l \\ tt}}^{(3)} + Q_{\substack{H q \\ tt}}^{(3)} \right] + \frac{g_2^2}{2} Q_V  +\frac{g_2^2}{2} \sum_{\psi = u,d,e}Y_\psi Q_{\psi H},
\nn
\cO_{2B} &=  -\frac{g_1^4}{2}\hyp_h^2  Q_{H \Box} - 2 g_{1}^4 \hyp_h^2 Q_{H D} - g_{1}^4 \hyp_h \left[\hyp_l Q_{\substack{H l \\ tt}}^{(1)} +\hyp_e Q_{\substack{H e \\ tt}} + \hyp_q Q_{\substack{H q \\ tt}}^{(1)}+\hyp_u Q_{\substack{H u \\ tt}}+ \hyp_d Q_{\substack{H d \\ tt}} \right] - \frac{g_1^4}{2}j_1^\mu j_{1\mu} \nn
\cO_{2W} &= -\frac{g_2^4}{4}\left[ Q_{\substack{H l \\ tt}}^{(3)} + Q_{\substack{H q \\ tt}}^{(3)}  +\frac{3}{2} Q_{H\Box} +Q_V  + \sum_{\psi = u,d,e}Y_\psi Q_{\psi H}\right] - \frac{g_2^4}{2} j_2^\mu j_{2\mu} \nn
\cO_{2G} &= -\frac{g_3^4}{2} j_3^\mu j_{3\mu} \nn
\cO_{HB}  &= \frac12 g_1^2 \hyp_h Q_{H \Box}  + 2g_1^2 \hyp_h Q_{H D} - \frac12 \hyp_h g_1^2 Q_{H B} - \frac14 g_1g_2  Q_{H WB} \nn
& +\frac12  g_1^2\left[\hyp_l Q_{\substack{H l \\ tt}}^{(1)} +\hyp_e Q_{\substack{H e \\ tt}} + \hyp_q Q_{\substack{H q \\ tt}}^{(1)}+\hyp_u Q_{\substack{H u \\ tt}}+ \hyp_d Q_{\substack{H d \\ tt}} \right] ,\nn
\cO_{HW} &= \frac34 g_2^2 Q_{H \Box} - \frac14 g_2^2 Q_{H W} - \frac12 \hyp_h g_1 g_2  Q_{H WB} + \frac14  g_2^2\left[Q_{H l}^{(3)} + Q_{H q}^{(3)} \right] 
\nn
&+ \frac{g_2^2}{2} Q_V +\frac{g_2^2}{2} \sum_{\psi = u,d,e}Y_\psi Q_{\psi H} \,,
\end{align}
as well as $\cO_{\gamma} = g_1^2 Q_{HB}$, $\cO_{g} = g_3^2 Q_{HG}$, $\cO_{3W} = g_2^3 Q_{W}$, and $\cO_{3G} = g_3^3 Q_{G}$, where the $Q_i$ are Warsaw basis operators. We have used the identity $(H^\dagger  \overleftrightarrow {D_\mu}^I H)^2 = 4\lambda v^2 (H^\dagger H)^2 + 6 \cO_H - 8 \cO_6 - 2\cO_y$ to reduce the square of Higgs triplet currents.  We have also defined 
\begin{align}
Q_V &= 4\lambda Q_H -2\lambda v^2 (H^\dagger H)^2 \,,\\
\cO_y &=\sum_{\psi = u,d,e}Y_\psi Q_{\psi H} = \left( [Y_u]_{rs} Q_{\substack{ uH \\ rs}} + [Y_d]_{rs} Q_{\substack{ dH \\ rs}}+ [Y_e]_{rs} Q_{\substack{ eH \\ rs}}+{\rm h.c.} \right) \,.
\end{align}
The squares of the SM currents in the Warsaw basis read
\begin{align}
j_1^\mu j_{1\mu} &=  \hyp_q^2 [Q_{qq}^{(1)}]_{pprr} + \hyp_l^2 [Q_{ll}]_{pprr}+ \hyp_u^2 [Q_{uu}]_{pprr} + + \hyp_d^2 [Q_{dd}]_{pprr} + + \hyp_e^2 [Q_{ee}]_{pprr} \nn
&+2\hyp_q \left( \hyp_l [Q_{lq}^{(1)}]_{pprr} + \hyp_u [Q_{qu}^{(1)}]_{pprr} + \hyp_d [Q_{qd}^{(1)}]_{pprr} + \hyp_e [Q_{qe}]_{pprr} \right) \nn
&+2\hyp_l \left( \hyp_u [Q_{lu}]_{pprr} + \hyp_d [Q_{ld}]_{pprr} + \hyp_e [Q_{le}]_{pprr} \right) \nn
&+2\hyp_u \left(\hyp_d [Q_{ud}^{(1)}]_{pprr} + \hyp_e [Q_{eu}]_{pprr} \right) + 2 \hyp_d \hyp_e [Q_{ed}]_{pprr} \,, \\
j_2^\mu j_{2\mu} &= \frac{1}{4}\left( [Q_{qq}^{(3)}]_{pprr} + 2[Q_{lq}^{(3)}]_{pprr} + 2[Q_{ll}]_{prrp} - [Q_{ll}]_{pprr} \right) \,,\\
j_3^\mu j_{3\mu} &= -\frac{1}{6} [Q_{qq}^{(1)}]_{pprr} + \frac{1}{4} [Q_{qq}^{(1)}]_{prrp} + \frac{1}{4} [Q_{qq}^{(3)}]_{prrp} -\frac{1}{6} [Q_{uu}]_{pprr} + \frac{1}{2} [Q_{uu}]_{prrp} \nn
&-\frac{1}{6} [Q_{dd}]_{pprr} +\frac{1}{2} [Q_{dd}]_{prrp} + 2[Q_{qu}^{(8)}]_{pprr} + 2[Q_{qd}^{(8)}]_{pprr} + + 2[Q_{ud}^{(8)}]_{pprr}\,,
\end{align}
where in $j_2^\mu j_{2\mu}$ we used the Fierz identity $(\bar l^p \gamma_{\mu} \tau^I l^p)(\bar l^r \gamma^{\mu} \tau^I l^r) = 2[Q_{ll}]_{prrp} - [Q_{ll}]_{pprr} $.
The non-universal contributions are already in the Warsaw basis, except for operators of the form $\cO_{\psi D}$. The map for those is
\begin{align}
O_{qD}^{(3)} & = \frac{g_2^2}{2} \left[Q_{\substack{H q \\ 33}}^{(3)} + Q_{\substack{qq \\ 33tt}}^{(3)} + Q_{\substack{lq \\ tt33}}^{(3)}\right] \,, \nn
O_{qD}^{(1)} & =  g_1^2 \left[\hyp_h Q_{\substack{H q \\ 33}}^{(1)} + \hyp_q Q_{\substack{qq \\ 33tt}}^{(1)} + \hyp_l Q_{\substack{lq \\ tt33}}^{(1)} + \hyp_u Q_{\substack{qu \\ 33tt}}^{(1)} + \hyp_d Q_{\substack{qd \\ 33tt}}^{(1)} + \hyp_e Q_{\substack{qe \\ 33tt}}\right] \,, \nn
O_{tD} & =  g_1^2 \left[\hyp_h Q_{\substack{H u \\ 33}} + \hyp_q Q_{\substack{qu \\ tt33}}^{(1)} + \hyp_l Q_{\substack{lu \\ tt33}} + \hyp_u Q_{\substack{uu \\ tt33}} + \hyp_d Q_{\substack{ud \\ 33tt}}^{(1)} + \hyp_e Q_{\substack{eu \\ tt33}}\right] \,.
\end{align}

\section{Complete effective operator bounds}
\label{app:B}
Here we report the bounds on all operators in~\cref{tab:UniversalOps,tab:nonUniversalOps,tab:UniversalOpsLoop}. We define the bound in either direction terms of the effective scales $\Lambda_{\pm} = \text{sign}(\cC_{\pm})|\cC_{\pm}|^{-1/2}$ and $\Lambda_{\rm bound} = \text{min}(|\Lambda_{-}|,|\Lambda_{+}|)$. In each direction, we also report the observable with the largest pull. In cases where $\Lambda_{\pm}$ has the ``wrong" sign for a given direction, that should be understood as a mild preference for NP at the corresponding scale. One can see that the scale we report for $\Lambda_{\rm bound}$ never corresponds to this choice.
\begin{table}
\centering
\begin{tabular}{|c|c|c|c|c|c|c|}
\hline
Wilson Coef. & $\Lambda_-$ [TeV] & [Obs]$_-$ & $\Lambda_+$ [TeV] & [Obs]$_+$ & [Obs]$_{\text{bound}}$ & $\Lambda_{\text{bound}}$ [TeV] \\
\hline
$\mathcal{C}_T$ & -45.52 & $m_W$ & 8.17 & $A_b^{\text{FB}}$ & $A_b^{\text{FB}}$ & 8.17 \\ \hline
$\mathcal{C}_{Hq}^{(1)}$ & 25.15 & $R_{\tau}$ & 3.98 & $R_{\tau}$ & $R_{\tau}$ & 3.98 \\ \hline
$\mathcal{C}_{Hq}^{(3)}$ & -10.89 & $R_b$ & 3.94 & $R_b$ & $R_b$ & 3.94 \\ \hline
$\mathcal{C}_{Ht}$ & -3.00 & $A_b^{\text{FB}}$ & 16.07 & $m_W$ & $A_b^{\text{FB}}$ & 3.00 \\ \hline
$\mathcal{C}_{Hq}^{(-)}$ & -9.32 & $m_W$ & 2.98 & $A_b^{\text{FB}}$ & $A_b^{\text{FB}}$ & 2.98 \\ \hline
$\mathcal{C}_B$ & -2.48 & $A_b^{\text{FB}}$ & 7.34 & $m_W$ & $A_b^{\text{FB}}$ & 2.48 \\ \hline
$\mathcal{C}_W$ & -2.41 & $A_b^{\text{FB}}$ & 6.85 & $m_W$ & $A_b^{\text{FB}}$ & 2.41 \\ \hline
$\mathcal{C}_{tW}$ & -5.63 & $m_W$ & 1.86 & $A_b^{\text{FB}}$ & $A_b^{\text{FB}}$ & 1.86 \\ \hline
$\mathcal{C}_{qD}^{(3)}$ & -3.68 & $R_b$ & 1.83 & $R_{\tau}$ & $R_{\tau}$ & 1.83 \\ \hline
$\mathcal{C}_{qq}^{(1)}$ & -1.50 & $R_{\tau}$ & -35.07 & $R_{\tau}$ & $R_{\tau}$ & 1.50 \\ \hline
$\mathcal{C}_{tB}$ & -4.34 & $m_W$ & 1.44 & $A_b^{\text{FB}}$ & $A_b^{\text{FB}}$ & 1.44 \\ \hline
$\mathcal{C}_{2W}$ & -3.61 & $m_W$ & 1.29 & $A_b^{\text{FB}}$ & $A_b^{\text{FB}}$ & 1.29 \\ \hline
$\mathcal{C}_{qt}^{(1)}$ & 5.59 & $R_{\tau}$ & 1.14 & $R_{\tau}$ & $R_{\tau}$ & 1.14 \\ \hline
$\mathcal{C}_{qD}^{(1)}$ & -12.43 & $m_W$ & 1.12 & $A_b^{\text{FB}}$ & $A_b^{\text{FB}}$ & 1.12 \\ \hline
$\mathcal{C}_{tt}$ & -1.05 & $A_b^{\text{FB}}$ & 5.30 & $m_W$ & $A_b^{\text{FB}}$ & 1.05 \\ \hline
$\mathcal{C}_{qq}^{(3)}$ & -2.56 & $R_b$ & 0.94 & $R_b$ & $R_b$ & 0.94 \\ \hline
$\mathcal{C}_{tD}$ & -0.93 & $A_b^{\text{FB}}$ & 2.56 & $m_W$ & $A_b^{\text{FB}}$ & 0.93 \\ \hline
$\mathcal{C}_{2B}$ & -1.67 & $m_W$ & 0.77 & $A_b^{\text{FB}}$ & $A_b^{\text{FB}}$ & 0.77 \\ \hline
$\mathcal{C}_{HW}$ & 3.46 & $A_b^{\text{FB}}$ & 0.53 & $A_b^{\text{FB}}$ & $A_b^{\text{FB}}$ & 0.53 \\ \hline
$\mathcal{C}_{2G}$ & -0.53 & $R_b$ & -18.45 & $A_b^{\text{FB}}$ & $R_b$ & 0.53 \\ \hline
$\mathcal{C}_H$ & -0.47 & $A_b^{\text{FB}}$ & 2.21 & $m_W$ & $A_b^{\text{FB}}$ & 0.47 \\ \hline
$\mathcal{C}_g$ & -0.47 & \text{$H\to bb$} & 0.47 & \text{$H\to bb$} & \text{$H\to bb$} & 0.47 \\ \hline
$\mathcal{C}_{3W}$ & -1.38 & $m_W$ & 0.46 & $A_b^{\text{FB}}$ & $A_b^{\text{FB}}$ & 0.46 \\ \hline
$\mathcal{C}_{tG}$ & -0.46 & $A_b^{\text{FB}}$ & 1.47 & $m_W$ & $A_b^{\text{FB}}$ & 0.46 \\ \hline
$\mathcal{C}_{HB}$ & -0.49 & $A_b^{\text{FB}}$ & 0.38 & $A_e$ [SLD] & $A_e$ [SLD] & 0.38 \\ \hline
$\mathcal{C}_{\gamma}$ & -1.00 & $m_W$ & 0.34 & $A_b^{\text{FB}}$ & $A_b^{\text{FB}}$ & 0.34 \\ \hline
$\mathcal{C}_{tH}$ & -0.18 & \text{$H\to \mu \mu$} & 0.27 & \text{$H \to \tau \tau$} & \text{$H \to \mu \mu$} & 0.18 \\ \hline
$\mathcal{C}_{qt}^{(8)}$ & 0.59 & $R_{\tau}$ & 0.11 & $R_{\tau}$ & $R_{\tau}$ & 0.11 \\ \hline
$\mathcal{C}_{3G}$ & -0.32 & $m_W$ & 0.10 & $A_b^{\text{FB}}$ & $A_b^{\text{FB}}$ & 0.10 \\ \hline
\end{tabular}
\caption{95\% CL single operator bounds all SILH+TQ operators (current EW precision data). We also show the bound on $\cC_{Hq}^{(-)}$, defined as the WC of the operator $\cO_{Hq}^{(-)} = \cO_{Hq}^{(1)}-\cO_{Hq}^{(3)}$. The Wilson coefficients are renormalized at $\Lambda = 2.5$ TeV.}
\label{tab:currentBoundsAll}
\end{table}

\begin{table}
\centering
\begin{tabular}{|c|c|c|c|c|c|c|}
\hline
Wilson Coef. & $\Lambda_{-}$ [TeV] & [Obs]$_{-}$ & $\Lambda_{+}$ [TeV] & [Obs]$_{+}$ & [Obs]$_{\text{bound}}$ & $\Lambda_{\text{bound}}$ [TeV] \\
\hline
$\mathcal{C}_T$ & -74.24 & $m_W$ & 74.24 & $m_W$ & $m_W$ & 74.24 \\ \hline
$\mathcal{C}_{Hq}^{(1)}$ & -39.82 & $m_W$ & 39.82 & $m_W$ & $m_W$ & 39.82 \\ \hline
$\mathcal{C}_{Ht}$ & -35.92 & $m_W$ & 35.92 & $m_W$ & $m_W$ & 35.92 \\ \hline
$\mathcal{C}_{Hq}^{(-)}$ & -33.97 & $m_W$ & 33.97 & $m_W$ & $m_W$ & 33.97 \\ \hline
$\mathcal{C}_{tW}$ & -26.19 & $A_e$ & 26.19 & $A_e$ & $A_e$ & 26.19 \\ \hline
$\mathcal{C}_B$ & -26.15 & $A_e$ & 26.15 & $A_e$ & $A_e$ & 26.15 \\ \hline
$\mathcal{C}_{Hq}^{(3)}$ & -24.81 & $R_{\mu}$ & 24.81 & $R_{\mu}$ & $R_{\mu}$ & 24.81 \\ \hline
$\mathcal{C}_W$ & -24.67 & $A_e$ & 24.67 & $A_e$ & $A_e$ & 24.67 \\ \hline
$\mathcal{C}_{tB}$ & -20.24 & $A_e$ & 20.24 & $A_e$ & $A_e$ & 20.24 \\ \hline
$\mathcal{C}_{qq}^{(1)}$ & -16.59 & $m_W$ & 16.59 & $m_W$ & $m_W$ & 16.59 \\ \hline
$\mathcal{C}_{tt}$ & -14.64 & $m_W$ & 14.64 & $m_W$ & $m_W$ & 14.64 \\ \hline
$\mathcal{C}_{qt}^{(1)}$ & -14.61 & $m_W$ & 14.61 & $m_W$ & $m_W$ & 14.61 \\ \hline
$\mathcal{C}_{qD}^{(1)}$ & -13.73 & $A_e$ & 13.73 & $A_e$ & $A_e$ & 13.73 \\ \hline
$\mathcal{C}_{tD}$ & -12.9 & $A_e$ & 12.9 & $A_e$ & $A_e$ & 12.9 \\ \hline
$\mathcal{C}_{2W}$ & -12.48 & $A_e$ & 12.48 & $A_e$ & $A_e$ & 12.48 \\ \hline
$\mathcal{C}_{qD}^{(3)}$ & -11.73 & $R_{\mu}$ & 11.73 & $R_{\mu}$ & $R_{\mu}$ & 11.73 \\ \hline
$\mathcal{C}_{2B}$ & -8.58 & $A_e$ & 8.58 & $A_e$ & $A_e$ & 8.58 \\ \hline
$\mathcal{C}_{qq}^{(3)}$ & -7.95 & $R_{\mu}$ & 7.95 & $R_{\mu}$ & $R_{\mu}$ & 7.95 \\ \hline
$\mathcal{C}_{tG}$ & -7.91 & $A_e$ & 7.91 & $A_e$ & $A_e$ & 7.91 \\ \hline
$\mathcal{C}_{HW}$ & -6.91 & $m_W$ & 6.91 & $m_W$ & $m_W$ & 6.91 \\ \hline
$\mathcal{C}_{3W}$ & -6.44 & $A_e$ & 6.44 & $A_e$ & $A_e$ & 6.44 \\ \hline
$\mathcal{C}_H$ & -6.03 & $m_W$ & 6.03 & $m_W$ & $m_W$ & 6.03 \\ \hline
$\mathcal{C}_{2G}$ & -5.85 & $m_W$ & 5.85 & $m_W$ & $m_W$ & 5.85 \\ \hline
$\mathcal{C}_{\gamma}$ & -4.95 & $A_e$ & 4.95 & $A_e$ & $A_e$ & 4.95 \\ \hline
$\mathcal{C}_{HB}$ & -4.65 & $A_e$ & 4.65 & $A_e$ & $A_e$ & 4.65 \\ \hline
$\mathcal{C}_g$ & -3.63 & \text{$H\to bb$} & 3.63 & \text{$H\to bb$} & \text{$H\to bb$} & 3.63 \\ \hline
$\mathcal{C}_{3G}$ & -1.91 & $A_e$ & 1.91 & $A_e$ & $A_e$ & 1.91 \\ \hline
$\mathcal{C}_{qt}^{(8)}$ & -1.61 & $m_W$ & 1.61 & $m_W$ & $m_W$ & 1.61 \\ \hline
$\mathcal{C}_{tH}$ & -0.95 & \text{$H\to \tau \tau$} & 0.95 & \text{$H\to \tau \tau$} & \text{$H \to \tau \tau$} & 0.95 \\ \hline
\end{tabular}
\caption{95\% CL single operator bounds all SILH+TQ operators (FCC-ee projection). We also show the bound on $\cC_{Hq}^{(-)}$, defined as the WC of the operator $\cO_{Hq}^{(-)} = \cO_{Hq}^{(1)}-\cO_{Hq}^{(3)}$. The Wilson coefficients are renormalized at $\Lambda = 25$ TeV.}
\label{tab:FCCBoundsAll}
\end{table}

\section{Projected improvements for FCC-ee}
The observable improvements that we use for our FCC-ee projection are given in~\cref{tab:FCCeePROJZpole,tab:FCCeePROJother}. These include $W$- and $Z$-pole observables, as well as Higgs decays.
\label{sec:appFCC}
   \begin{table}[h]
 \centering
\renewcommand{\arraystretch}{0.9} 
 \begin{tabular}{c|c|c|c}
Observable & Current Rel. Error ($10^{-3}$) & FCC-ee Rel. Error ($10^{-3}$) & Proj. Error Reduction \\
\hline
\hline 
$ \mathrm{  \Gamma_Z} $ &  2.3 & 0.1 & 23 \\
\hline
$ \sigma_{\rm had}^0$ &  37 & 5 & 7.4  \\
$ R_b $ &  3.06 & 0.3 & 10.2    \\
$ R_c $ &  17.4 & 1.5 & 11.6    \\
$ A_{\rm FB}^{0,b} $ &  15.5 & 1 & 15.5    \\
$ A_{\rm FB}^{0,c} $ &  47.5 & 3.08 & 15.4    \\
$ A_b $ &  21.4 & 3 & 7.13    \\
$ A_c $ &  40.4 & 8 & 5.05    \\
\hline
$ R_e $ &  2.41 & 0.3 & 8.03    \\
$ R_\mu $ &  1.59 & 0.05 & 31.8    \\
$ R_\tau $ &  2.17 & 0.1 & 21.7    \\
$ A_{\rm FB}^{0,e} $ &  154 & 5 & 30.8    \\
$ A_{\rm FB}^{0,\mu} $ &  80.1 & 3 & 26.7    \\
$ A_{\rm FB}^{0,\tau} $ &  104.8 & 5 & 21    \\
$ A_e^{**}$ &  12.5 & 0.13 & 95    \\
$ A_\mu^{**} $ &  102 & 0.15 & 680    \\
$ A_\tau^{**} $ &  102 & 0.3 & 340    \\
\hline
 \end{tabular}
 \caption{Projected FCC-ee improvement for $Z$-pole observables from~\cite{DeBlas:2019qco}. The $A_\ell^{**}$ are from lepton
polarization and LR asymmetry measurements at SLC. For $ A_e^{**}$, we use the projection in~\cite{deBlas:2022ofj}. }
 \label{tab:FCCeePROJZpole}
 \end{table}

\begin{table}[h]
 \centering
\renewcommand{\arraystretch}{1.} 
 \begin{tabular}{c|c|c|c|c}
Observable & Value  & Error  & FCC-ee Tot. & Proj. Error Red. \\
\hline
\hline 
$ \mathrm{  \Gamma_W  ~(MeV) } $ &  2085 & 42 & 1.24 & 34 \\
$ \mathrm{  m_W  ~(MeV) } $ &  80350 & 15 & 0.39 & 38 \\
$ \tau \rightarrow \mu \nu\nu (\%)$ &  17.38 & 0.04 & 0.003 & 13 \\
$ {\rm Br}(W\rightarrow e \nu) (\%)$ & 10.71 & 0.16 & 0.0032 & 50 \\
$ {\rm Br}(W\rightarrow \mu \nu) (\%)$ & 10.63 & 0.15 & 0.0032 & 47 \\
$ {\rm Br}(W\rightarrow \tau \nu) (\%)$ & 11.38 & 0.21 & 0.0046 & 46 \\
$ \mu_{ b\bar b}$ &  0.99 & 0.12 & 0.003 & 40 \\
$ \mu_{c\bar c}$ &  8 & 22 & 0.022 & 1000 \\
$ \mu_{\tau\bar \tau} $ & 0.91 & 0.09 & 0.009 & 10 \\
$ \mu_{\mu \bar \mu} $ & 1.21 & 0.35 & 0.19 & 1.84 \\
\hline
 \end{tabular}
 \caption{Projected FCC-ee improvement for selected $H$, $\tau$ and $W$-pole observables from~\cite{DeBlas:2019qco,Blondel:2021ema,Bernardi:2022hny}.}
 \label{tab:FCCeePROJother}
 \end{table}

\section{Additional plots}
\label{app:C}
\begin{figure}[h]%
    \centering
    \subfloat[\centering Left compositeness]{{\includegraphics[width=0.31\textwidth]{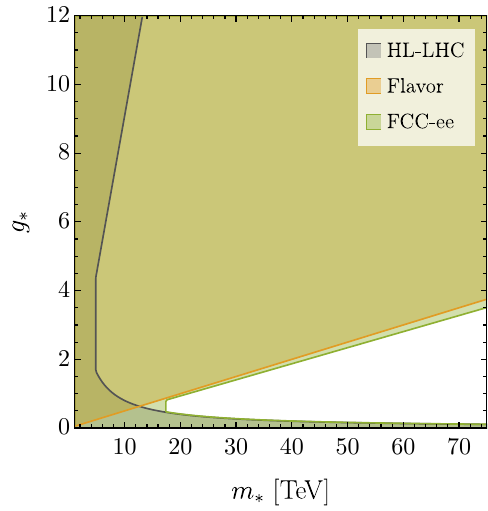} }}%
    \hspace{1mm}
    \subfloat[\centering Mixed compositeness]{{\includegraphics[width=0.31\textwidth]{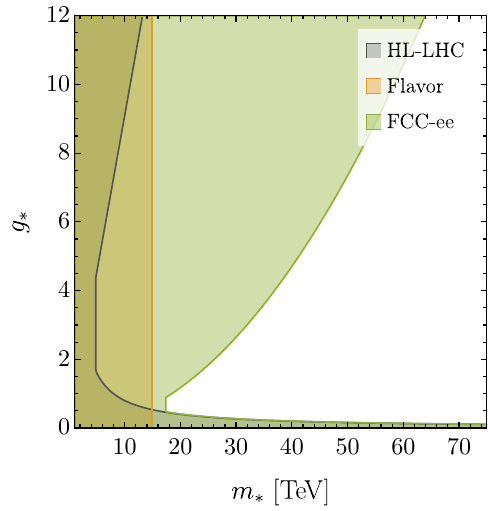} }}%
    \hspace{1mm}
    \subfloat[\centering Right compositeness]{{\includegraphics[width=0.31\textwidth]{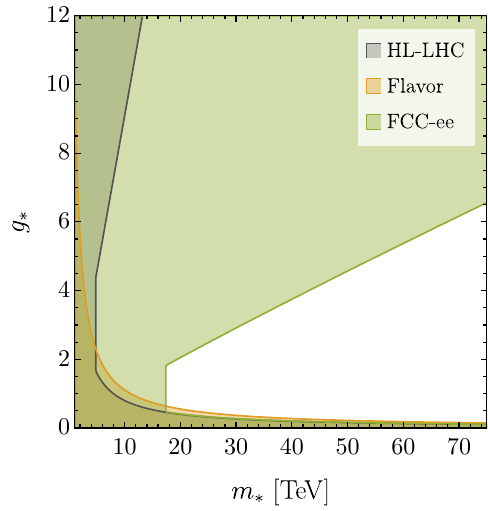} }}%
    \caption{Summary plot of future projections by sector, assuming limitations due to intrinsic theoretical uncertainties (see the text of~\cref{app:C} for details).  We assume $P_{LR}$ symmetry for the FCC-ee projection, and $\Lambda = $ 25 TeV. The white region is allowed.}%
    \label{fig:summaryPlotTL}%
    \vspace{5mm}
\end{figure}
\begin{figure}[h]
    \centering
    \subfloat[\centering Left compositeness (current)]{{\includegraphics[width=0.485\textwidth]{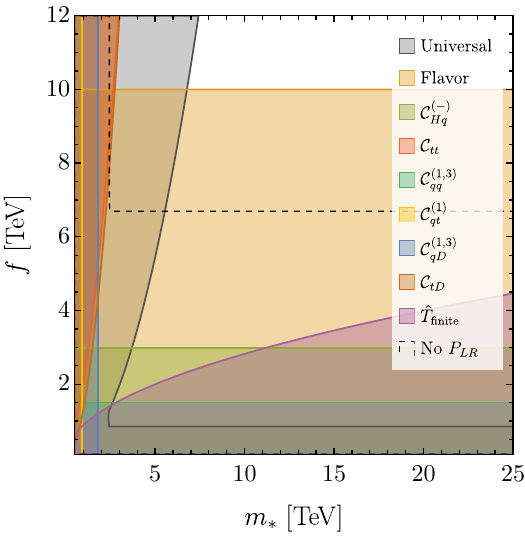} }}%
    \hspace{2.35mm}
    \subfloat[\centering Left compositeness (FCC-ee)]{{\includegraphics[width=0.475\textwidth]{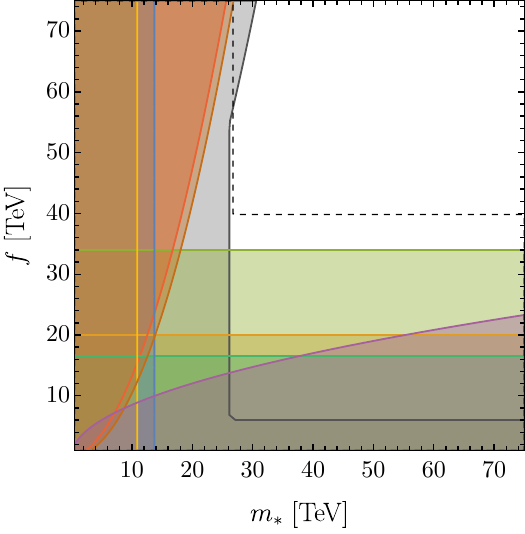} }}%
    \caption{Left compositeness scenario, same as~\cref{fig:LeftCompPlots} except we traded $g_*$ for $f$. The legend is the same for both panels, and the white region is allowed.}%
    \label{fig:LeftCompNatPlots}%
\end{figure}
\begin{figure}[h]
    \centering
    \subfloat[\centering Mixed compositeness (current)]{{\includegraphics[width=0.485\textwidth]{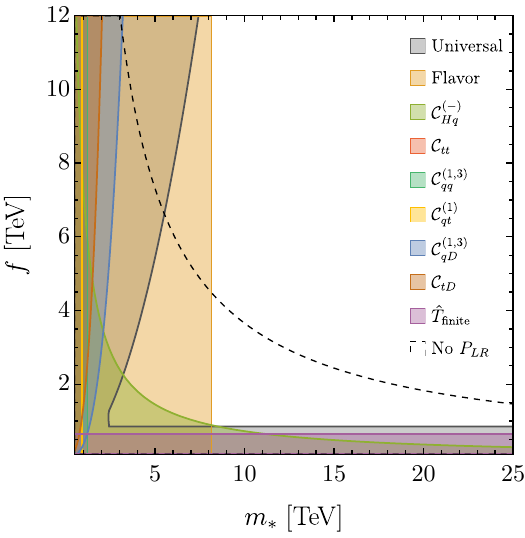} }}%
    \hspace{2.35mm}
    \subfloat[\centering Mixed compositeness (FCC-ee)]{{\includegraphics[width=0.475\textwidth]{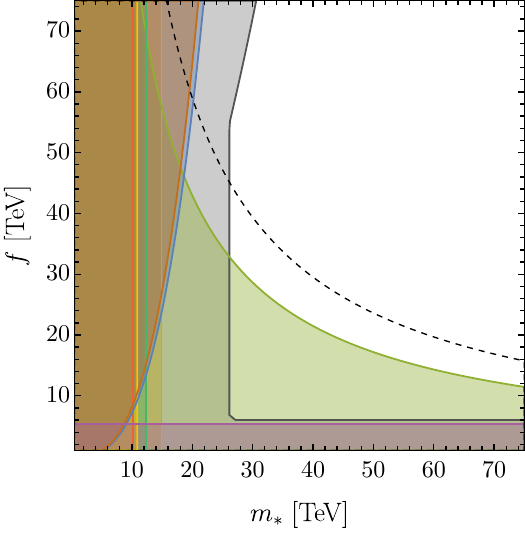} }}%
    \caption{Mixed compositeness scenario, same as~\cref{fig:MixedCompPlots} except we traded $g_*$ for $f$. The legend is the same for both panels, and the white region is allowed.}%
    \label{fig:MixedCompNatPlots}%
\end{figure}
\begin{figure}[h]
    \centering
    \subfloat[\centering Right compositeness (current)]{{\includegraphics[width=0.485\textwidth]{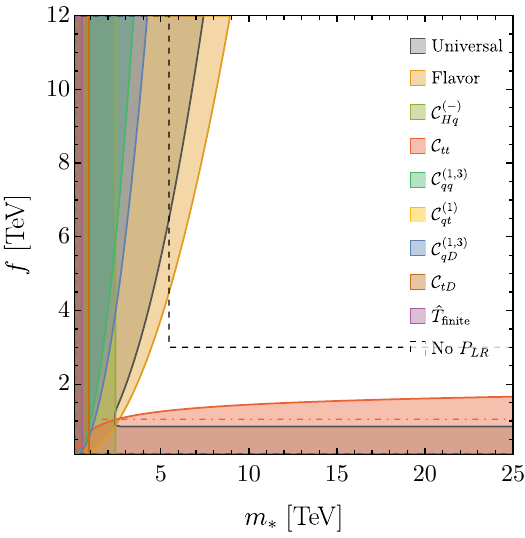} }}%
    \hspace{2.35mm}
    \subfloat[\centering Right compositeness (FCC-ee)]{{\includegraphics[width=0.475\textwidth]{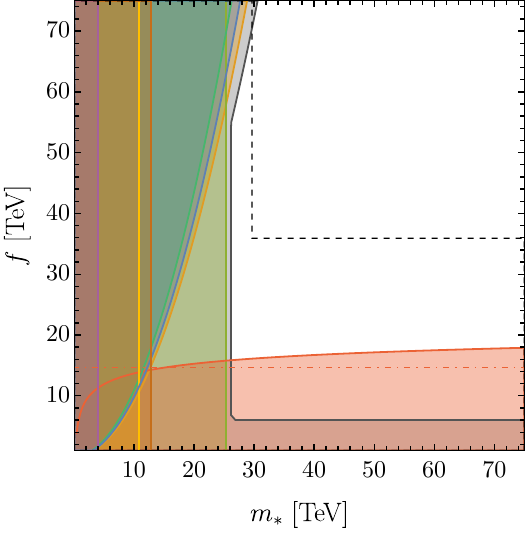} }}%
    \caption{Right compositeness scenario, same as~\cref{fig:RightCompPlots} except we traded $g_*$ for $f$. The legend is the same for both panels, and the white region is allowed. The solid vs. dot-dashed red line is explained in~\S~\ref{sec:Rcomp}}%
    \label{fig:RightCompNatPlots}%
\end{figure}

Here we provide some additional plots to complement the information given in the main text. In~\cref{fig:LeftCompNatPlots,fig:MixedCompNatPlots,fig:RightCompNatPlots}, we show the same results displayed in~\cref{fig:LeftCompPlots,fig:MixedCompPlots,fig:RightCompPlots}, but we have exchanged $g_*$ for the scale $f = m_*/g_*$ on the $y$-axis. Generally speaking, the lower $f$ is allowed to be, the more natural the model is.

Finally,~\cref{fig:summaryPlotTL} shows the same as~\cref{fig:summaryPlot}, but assuming some intrinsic theoretical uncertainty on the measurements which dominate our bounds, namely $m_W$ and $A_e$. The dominant source of theoretical error on universal contributions to the latter is the uncertainty on the effective Weinberg angle $\sin^2\theta_W^{\rm eff}$, which contains the net contributions from all radiative corrections. We assume an intrinsic theory uncertainty of $[\Delta \sin^2\theta_W^{\rm eff}]_{\rm Th} \simeq 10^{-5}$~\cite{Blondel:2018mad,Freitas:2019bre}. Thus, in making~\cref{fig:summaryPlotTL}, we use
\begin{align}
[\Delta m_W]_{\rm Th} &\simeq 1~{\rm MeV} \,,\\
[\Delta A_\ell]_{\rm Th} &\simeq 5 \times 10^{-5} \,.
\end{align}
These absolute theoretical uncertainties are about a factor of 3 (2) larger than the projected absolute experimental error in the case of $m_W$ ($A_e$). The bounds in the $(m_*,g_*)$ plane therefore change by roughly the square root of these factors, as can be seen in~\cref{fig:summaryPlotTL}.

\FloatBarrier
\bibliographystyle{JHEP}
\bibliography{refs}

\end{document}